# The Extreme Space Weather Event of 1872 February: Sunspots, Magnetic Disturbance, and Auroral Displays

Hisashi Hayakawa[1,2], Edward W. Cliver[3], Frédéric Clette[4], Yusuke Ebihara[5,6], Shin Toriumi[7], Ilaria Ermolli[8], Theodosios Chatzistergos[9], Kentaro Hattori[10], Delores J. Knipp[11], Séan P. Blake[12,13], Gianna Cauzzi[3], Kevin Reardon[3], Philippe-A. Bourdin[14,15], Dorothea Just[15], Mikhail Vokhmyanin[16], Keitaro Matsumoto[1], Yoshizumi Miyoshi[1], José R. Ribeiro[17], Ana P. Correia[17], David M. Willis[2], Matthew N. Wild[2], and Sam M. Silverman[18]

[1] Institute for Space-Earth Environmental Research and Institute for Advanced Researches, Nagoya University, Nagoya 4648601, Japan; hisashi@nagoya-u.jp, hisashi.hayakawa@stfc.ac.uk
[2] Science and Technology Facilities Council, RAL Space, Rutherford Appleton Laboratory, Harwell Campus, Didcot OX11 0QX, UK
[3] National Solar Observatory, Boulder, CO 80303, USA; ecliver@nso.edu
[4] World Data Center SILSO, Observatoire Royal de Belgique, Brussels 1180, Belgium
[5] Research Institute for Sustainable Humanosphere, Kyoto University, Uji 6100011, Japan
[6] Unit of Synergetic Studies for Space, Kyoto University, Kyoto 6068306, Japan
[7] Institute of Space and Astronautical Science, Japan Aerospace Exploration Agency, Sagamihara, Kanagawa 2525210, Japan
[8] INAF, Osservatorio Astronomico di Roma, Monte Porzio Catone I-00078, Italy
[9] Max Planck Institute for Solar System Research, Göttingen D-37077, Germany
[10] Faculty of Societal Safety Sciences, Kansai University, Takatsuki 5691116, Japan
[11] Smead Aerospace Engineering Sciences Department 429 UCB, University of Colorado Boulder, Boulder, CO 80309, USA
[12] Heliophysics Science Division, NASA Goddard Space Flight Center, Greenbelt, MD, USA
[13] Catholic University of America, Washington, DC, USA
[14] University of Graz, Institute of Physics, Universitätsplatz 5, 8010 Graz, Austria
[15] Institut für Weltraumforschung, Österreichische Akademie der Wissenschaften, 8042 Graz, Austria
[16] University of Oulu, Oulu FI-90014, Finland
[17] Escola Secundária Henrique Medina, Esposende Av. Dr. Henrique Barros Lima 4740-203 Esposende, Portugal
[18] 18 Ingleside Road, Lexington, MA 02420, USA
Received 2023 February 3; revised 2023 March 9; accepted 2023 March 22; published 2023 December 1

## Abstract

We review observations of solar activity, geomagnetic variation, and auroral visibility for the extreme geomagnetic storm on 1872 February 4. The extreme storm (referred to here as the Chapman–Silverman storm) apparently originated from a complex active region of moderate area ($\approx 500~\mu$sh) that was favorably situated near disk center (S19° E05°). There is circumstantial evidence for an eruption from this region at 9–10 UT on 1872 February 3, based on the location, complexity, and evolution of the region, and on reports of prominence activations, which yields a plausible transit time of $\approx 29$ hr to Earth. Magnetograms show that the storm began with a sudden commencement at $\approx 14$:27 UT and allow a minimum Dst estimate of $\leq -834$ nT. Overhead aurorae were credibly reported at Jacobabad (British India) and Shanghai (China), both at 19°.9 in magnetic latitude (MLAT) and 24°.2 in invariant latitude (ILAT). Auroral visibility was reported from 13 locations with MLAT below $|20|°$ for the 1872 storm (ranging from $|10°.0|$–$|19°.9|$ MLAT) versus one each for the 1859 storm ($|17°.3|$ MLAT) and the 1921 storm ($|16°.2|$ MLAT). The auroral extension and conservative storm intensity indicate a magnetic storm of comparable strength to the extreme storms of 1859 September (25°.1 ± 0°.5 ILAT and $-949 \pm 31$ nT) and 1921 May (27°.1 ILAT and $-907 \pm 132$ nT), which places the 1872 storm among the three largest magnetic storms yet observed.

*Unified Astronomy Thesaurus concepts:* Solar active regions (1974); Solar flares (1496); Solar coronal mass ejections (310); Aurorae (2192); Geomagnetic fields (646); Magnetic storms (2320)

## 1. Introduction

Our civilization has become increasingly vulnerable to extreme solar storms because of our growing dependency on technology-based infrastructure (Pulkkinen 2007; Baker et al. 2008; Beggan et al. 2013; Baker & Lanzerotti 2016; Knipp et al. 2016; Lanzerotti 2017; Pulkkinen et al. 2017; Oughton et al. 2019; Hapgood et al. 2021). Therefore, it is essential to document both the solar origins (Temmer 2021) and the terrestrial impacts of extreme space weather events (Pulkkinen 2007). In particular, the solar-induced magnetic storms pose a systemic threat to the electrical power grids. However, it is challenging to study such extreme space weather events because of their low occurrence (Riley & Love 2017; Usoskin 2017; Gopalswamy 2018; Riley et al. 2018; Miyake et al. 2019; Chapman et al. 2020a, 2020b; Cliver et al. 2022a) and the short chronological coverage of the modern databases that generally began with the International Geophysical Year in 1957–1958 (Lanzerotti & Baker 2018; Lanzerotti 2017; Hayakawa et al. 2023b). To date, the best observed and documented extreme magnetic storm (in terms of solar origin, interplanetary disturbances, geomagnetic impact, and the resultant auroral extension) is that of 1989 March 13–14, which was the most intense storm in the space age (Allen et al. 1989; Odenwald 2007; Lakhina et al. 2013; Cid et al. 2014; Saiz et al. 2016; Boteler 2019). The limited chronological coverage of the modern database requires us to go back in time to historical archives to investigate comparable (or greater) storms before the space age.

Sustained systematic magnetic observations have been conducted since the 1830s (Chapman & Bartels 1940; Cawood 1979; Honigmann 1984; Beggan et al. 2023). Within this coverage, the Carrington event of 1859 September is considered to be an exemplar for extreme space weather events in terms of solar flare magnitude, solar wind speed, magnetic storm intensity, and equatorward extension of the auroral visibility (Tsurutani et al. 2003; Cliver & Svalgaard 2004; Nevanlinna 2008; Cliver & Dietrich 2013; Lakhina & Tsurutani 2016; Hayakawa et al. 2016, 2020; Hudson 2021; Cliver et al. 2022a). This event has been regarded as one of the worst-case scenarios in numerous space weather studies (Daglis 2000, 2004; Baker et al. 2008; Hapgood 2011; Riley 2012; Vennerstrøm et al. 2016; Riley & Love 2017; Riley et al. 2018; Dyer et al. 2018; Oughton et al. 2019; Hapgood et al. 2021). However, recent studies have begun to cast doubt on its pre-eminence in terms of magnetic storm intensity. For example, the extreme interplanetary coronal mass ejection (ICME) of 2012 July, which missed the Earth, was shown to have the potential to have caused an even more extreme magnetic storm than the Carrington event if it had hit the Earth at an optimal universal time (UT) and season for solar wind-magnetospheric coupling (Russell & McPherron 1973; Cliver et al. 2000; Baker et al. 2013; Russell et al. 2013; Liu et al. 2014, 2019).

Archival investigations of the historical magnetic measurements allow us to extend quantitative analyses for great magnetic storms beyond the International Geophysical Year (Mursula et al. 2008; Lockwood et al. 2018; Love et al. 2019a, 2019b; Hayakawa et al. 2023b). These studies have estimated a minimum disturbance-storm-time (Dst) index (Sugiura 1964; Sugiura & Kamei 1991) for the extreme storm of 1921 May of $\approx -907 \pm 132$ nT (Love et al. 2019b), comparable to the disturbance variation for the Carrington storm at Bombay of $\approx -949 \pm 31$ nT (Hayakawa et al. 2022a; see also Siscoe et al. 2006; Cliver & Dietrich 2013). Furthermore, surveys of historical auroral records allow us to extend the space weather chronology for millennia (Hattori et al. 2019; Carrasco & Vaquero 2020; Knipp et al. 2021) and show significant extensions of the equatorward boundary of auroral oval during the storms in 1770 September, 1872 February, and 1921 May to be comparable to or latitudinally lower than that of the 1859 September storm (Chapman 1957a; Silverman & Cliver 2001; Silverman 2006; Hayakawa et al. 2019b; Hapgood 2019; Hayakawa et al. 2020).

In this context, the February 1872 storm falls into the category of extreme geomagnetic storms (Chapman 1957a; Silverman 2008; Hayakawa et al. 2019a; Cliver et al. 2022a). Chapman (1957a) listed this event in the category of "outstanding auroras" alongside the Carrington storm based on the auroral visibility at Bombay. Silverman (2008) considered this event to be even more extreme than the Carrington storm (Kimball 1960; Tsurutani et al. 2003; Silverman 2006) after having reconstructed its equatorward boundary of auroral visibility. Hereafter, we refer to the 1872 February storm as the Chapman–Silverman (C-S) storm. Silverman (1995) first suggested that the report of an aurora at Bombay (10°.0 magnetic latitude (MLAT) in 1872) was due to conflation with reported telegraph outages, whereas upon further consideration Silverman (2008) accepted the report of the sighting at Bombay as credible—but as sporadic aurora. Hayakawa et al. (2018) extended archival investigations for historical auroral records in East Asia beyond what had been initially cataloged in Willis et al. (2007) and reconstructed the equatorward boundary of the auroral oval in the East Asian sector as $\approx 24°$ invariant latitude (ILAT). Hayakawa et al. (2018) determined that the auroral visibility from Bombay was geometrically possible in terms of triangulation in reference to confirmed overhead aurorae observed at $\approx 20°$ MLAT in East Asia. Valach et al. (2019) and Berrilli & Giovannelli (2022) have investigated magnetic measurements in Greenwich and Rome, respectively, for the 1872 February storm to compare them with the local auroral behavior in the European sector. However, a comprehensive review of this storm—one that considers in detail its solar origin, lowest latitude of overhead approach at all night-time longitudes, and temporal and spatial relationship of geomagnetic and auroral activity—has not yet been undertaken. We do so here by combining existing knowledge of the Chapman–Silverman storm with newly uncovered solar, geomagnetic, and auroral observations.

In Section 2, we review contemporary observations of sunspots and other solar features in early February of 1872 to better define the source of the extreme storm. In Section 3, we present magnetic measurements from Greenwich, Colaba, Tiflis, and elsewhere for the 1872 February storm to draw a comparison with the associated auroral records. In Section 4, we compile and analyze the auroral observations, including recently uncovered reports from the Middle East, the Indian Ocean, and Africa for the Chapman–Silverman storm, and compare their temporal evolutions in each geographical sector with geomagnetic measurements. Reports of low-latitude aurora at $<|20|°$ MLAT are discussed in Section 5. The results are summarized and discussed in Section 6.

## 2. Solar Activity

### 2.1. Sunspot Number and Solar Disk Features in 1872 January–March

The Chapman–Silverman storm of 1872 February 4 occurred in the declining phase of Solar Cycle 11, which had its maximum in 1870 August (Hathaway 2015; Clette & Lefèvre 2016). Empirically, intense space weather events frequently occur during the declining phase of large solar cycles (Lefèvre et al. 2016; Owens et al. 2022). In the week preceding the event, the international sunspot number (Version 2.0; Clette et al. 2014, 2015; Clette & Lefèvre 2016) increased significantly from a value of 95 on January 28 to 272 on February 2, following the appearance on the solar disk of the active region that we identify as the source of the extreme storm, as shown in Figure 1. Archival records available at the libraries of the INAF Osservatorio Astronomico di Roma and Osservatorio Astrofisico di Catania document detailed solar observations by Angelo Secchi (1873, pp. 8–12 and 16–19) in Rome (see Secchi 1872b, p. 83; MS OAR 1–4),[19] Pietro Tacchini (1872, p. 83) in Palermo, and Francesco Denza in Moncalieri (Secchi 1872b, p. 83) during 1872 (Figure 1; see footnote 19). Secchi's sunspot drawing on January 27 shows a significant increase of prominence activity on the solar limb immediately before the entrance of the group numbered 29 in

---

[19] So far, Secchi's sunspot observations in 1872 January–November and Denza's sunspot observations have been largely overlooked, even in the latest database for sunspot group number (Vaquero et al. 2016). These records will be of great benefit in further revisions and recalibrations of the sunspot group number and relative sunspot number (Clette et al. 2023).

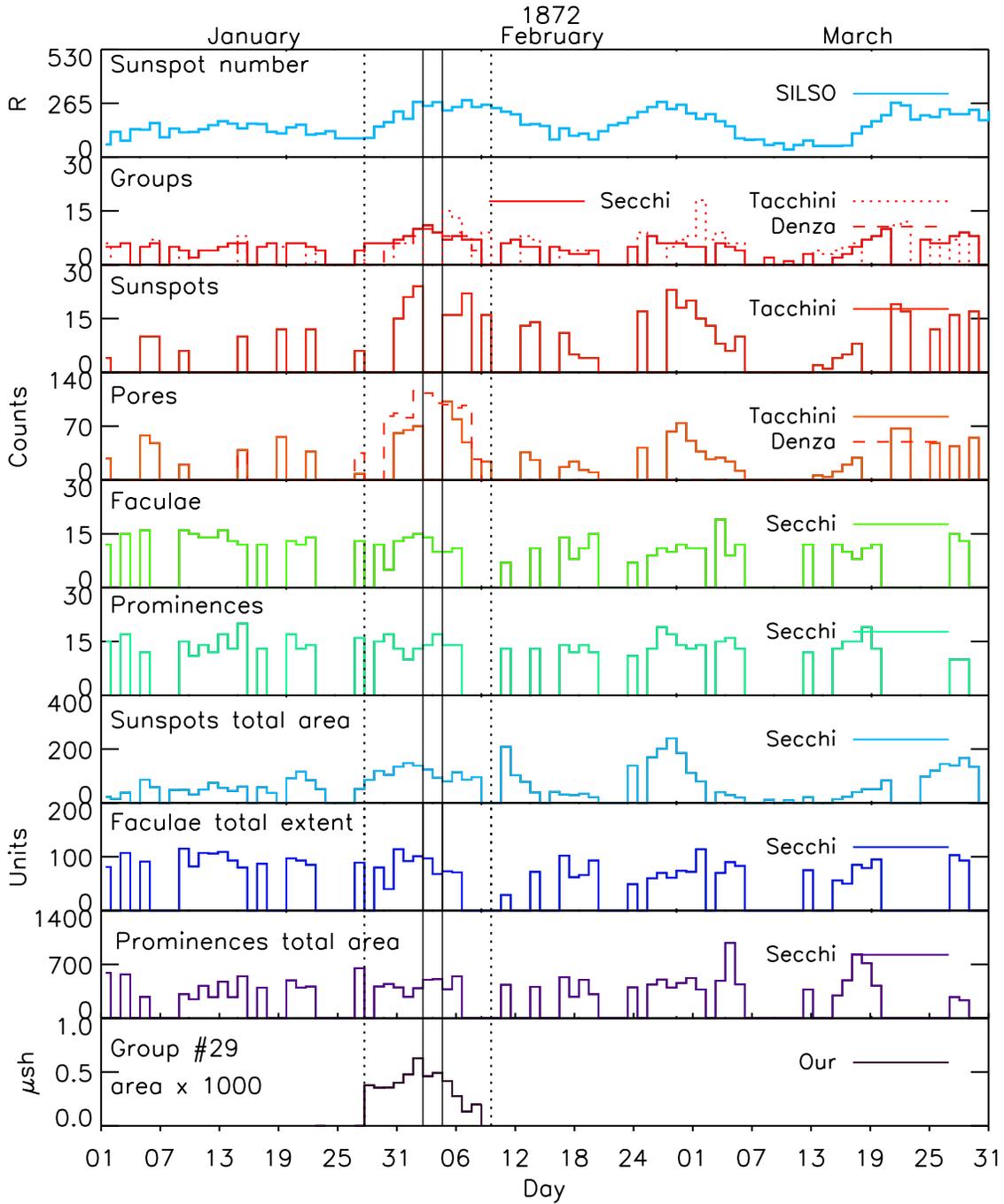

**Figure 1.** Solar activity around the time of the Chapman–Silverman storm: The top panel shows the daily international sunspot number (Version 2; Clette & Lefèvre 2016) time series maintained by the World Data Centre sunspot Index and Long-term Solar Observations (Source: World Data Centre SILSO). The other panels show counts of sunspot groups, individual spots, pores, faculae, and prominences, as well as corrected areas of sunspots, faculae, and prominences derived from archival documents from the solar observations of Angelo Secchi, Pietro Tacchini, and Francesco Denza. The bottom panel shows the evolution of the fractional corrected area for Secchi's sunspot Group #29 in millionths of a solar hemisphere ($\mu$sh). Secchi's units of feature extents are given in 1mm × 1mm (width × height), 1° (width), and 1° × 1mm (width × height), which correspond to 8 × 8 arcsec$^2$, 16.38 arcsec, 16.38 × 8 arcsec$^2$, for sunspots, faculae, and prominences, respectively. Note that the total linear extent of the faculae is reported, not their total area. Here, the areas in the last panel have been multiplied by 1000. The dotted-vertical lines indicate the interval (January 28 to February 9) during which Secchi's Spot Group #29 was on the solar disk. The vertical solid lines indicate 1872 February 3–4, which was the date of the storm.

Secchi's drawings (hereafter Group #29;[20] see Figure 2) with his annotation "(solar) activity has re-started (È ricominciata l'attività)" (Osservatorio Astronomico di Roma, MS INAF 1, f. 21) and associates this enhanced prominence activity with Group #29 (Secchi 1872c, p. 16).

### 2.2. Active Regions

Because the structure and complexity of solar active regions are essential keys to understanding the generation of solar eruptions (Toriumi & Wang 2019), we have analyzed the series of full-disk solar drawings by Angelo Secchi[21] (Figure 2), Pietro Tacchini[22] (Figure 3), and Gustave L. Bernaerts[23] (Figure 4) from their observations of the Sun in the weeks around the event. These three figures show the westward migration of the solar features to the left in the figures because the observations were performed with the telescopes in projection mode (mirror inversion in the east–west direction). The series of drawings of the three observers show the first appearance of sunspot Group #29 on January 28, in association with an increase in the sunspot number from 95 to 150 at that time (shown in Figure 1). These drawings also reveal a second sharp increase of the sunspot number on February 1–2 (from 188 to 272), which was mainly associated with an intrinsic growth of Group #29, whereas two small groups also came into sight at the east limb of the solar disk (see also Denza 1872, p. 825). Group #29 in Secchi's drawings reached its maximum extension on February 2 (627 $\mu$sh). On February 3, it started to shrink, with an area of 461 $\mu$sh; it was last seen near the west limb on February 9 by Bernaerts. Group #29 was thus not exceptionally large, similar to the medium-sized active region ($\approx$600 $\mu$sh) associated with a great magnetic storm (minimum Dst$^*$ $\approx$ −595 nT; Love et al. 2019a) on 1909 September 25 (Hayakawa et al. 2019a). The next largest groups on February 3 (Groups #28 and #34 in Figure 2) had areas of 166 $\mu$sh and 111 $\mu$sh, respectively.

---

[20] This group is labeled with the letter A, and tags z, z2, and z3 in the sunspot drawings of Tacchini and Bernaerts, respectively.

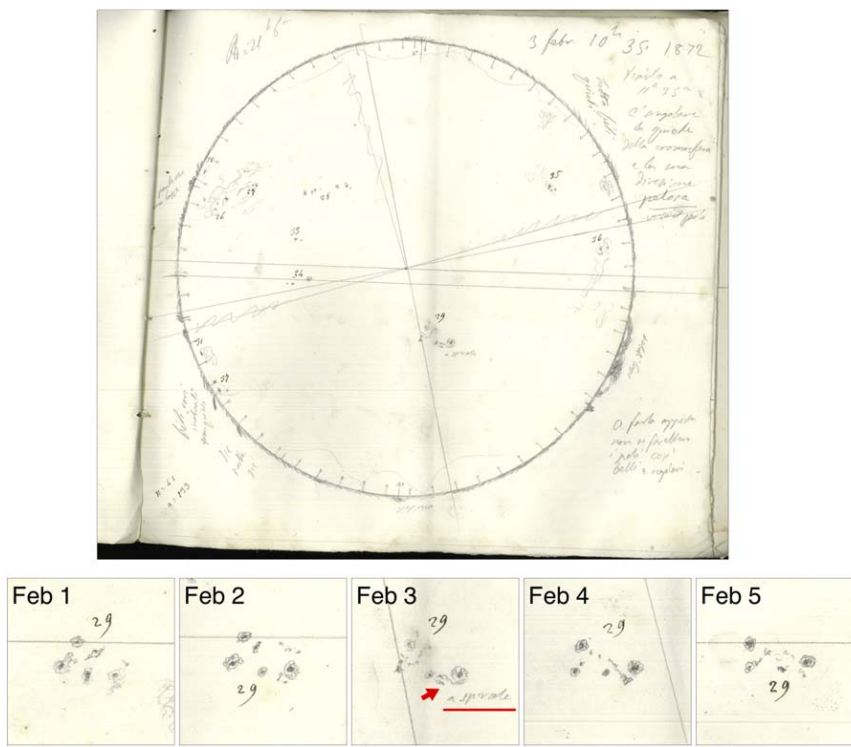

**Figure 2.** Top: Secchi's sunspot drawing on 1872 February 3 (Source: Osservatorio Astronomico di Roma, MS INAF 1, f. 28). Secchi annotated the Group #29 "in a spiral (*a spirale*)" on February 3. Bottom: Sequence of detailed sections from Secchi's drawings made over the period 1872 February 1–5 retracing the evolution of Group #29 (Source: Osservatorio Astronomico di Roma MS INAF 1, ff. 26—30). The east–west direction is inverted in these drawings because they were obtained by projection.

The "large group" in Meldrum (1872a) was almost certainly Secchi's Spot Group #29 in the southern hemisphere, and the "chain of sunspots" that Meldrum (1872a) mentioned corresponds to Spot Groups numbered 28, 33, 34, 27, 26, and 32 by Secchi in the northern hemisphere. Because the major magnetic storm occurred on February 4, the responsible flare and eruption likely occurred 1–2 days before the storm (i.e., February 2–3). Although Jones (1955) did not specify the source sunspot group for this storm, Group #29 seems to be the most plausible origin of the source flare because of its favorable location near the central meridian (S19° E05°) and its relatively large size (461 $\mu$sh) on February 3. Empirically, it has long been known that intense magnetic storms originate preferentially from regions in the vicinity of the central meridian (Newton 1943; Akasofu & Yoshida 1967; Cliver et al. 1990; Gopalswamy et al. 2012; Lefèvre et al. 2016; Cliver et al. 2022b).

Group #29 had a complex and unusual shape. No clear pair of dominant spots (global magnetic dipole) can be discerned. Contrary to most sunspot groups, which are elongated in the East-West direction (dipole parallel to the solar equator), this group did not have a particular axis of symmetry. The general shape and proximity of multiple individual sunspots suggest magnetic polarity mixing and a convoluted magnetic neutral line topology inside the group, and thus greater non-potential or free energy in this active region (Zirin & Liggett 1982; Török & Kliem 2003; Toriumi et al. 2017; Bourdin & Brandenburg 2018; Bourdin et al. 2018). At least three of the spots had substantial penumbrae on February 3, which indicates that they contained a strong magnetic flux.

On February 3, Secchi added an annotation, "in a spiral" (*a spirale*) to this group. This spiral structure may be related to the series of new pores extending from this spot (see the bottom of Figure 2 for its evolution from February 1–5), as seen in the satellite sunspot type of Toriumi et al. (2017) and Toriumi & Takasao (2017). If this is the case, then we can find a comparable pertinent example: NOAA active region (AR) 5395 of 1989 March (Figure 1 of Wang et al. 1991), where new sunspots (e.g., P4, P5, and P6) surrounded the large existing spot (F1) and produced a clockwise outward spiraling field or peculiar vortex-like motion of satellite sunspots (Tang & Wang 1993; Ishii et al. 1998). This peculiar configuration led to very powerful X-class flares (X15.0 on March 6 and X4.5 on March 10) and to a great magnetic storm with minimum Dst = −589 nT on March 14 (Allen et al. 1989; Tang & Wang 1993; WDC Kyoto et al. 2015).

---

[21] Osservatorio Astronomico di Roma MS INAF 1 (Disegni proiezioni macchie solari, Anno 1872).
[22] Osservatorio Astrofisico di Catania MS INAF 2 (Protuberanze, Spettri, Macchie dal 1 gennaio 1872 al 2 febbraio 1872).
[23] MS Bernaerts, v.3, ff. 25–26 in the Royal Astronomical Society.

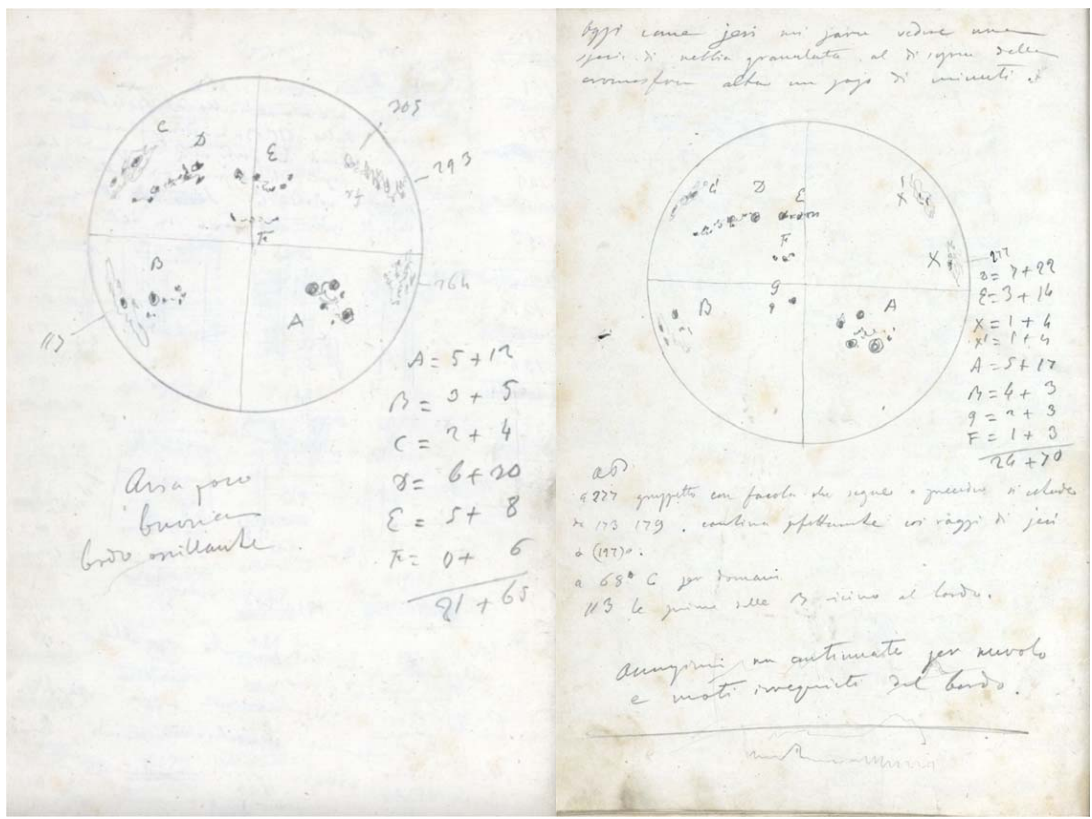

**Figure 3.** Tacchini's full-disk drawings made on 1872 February 1–2 (Source: Osservatorio Astrofisico di Catania, MS INAF 2, ff. 34 and 36).

In summary, Group #29 in Secchi's drawings was most likely the source of the C-S storm, based on its rapid evolution, favorable location, relative size, and complexity.[24]

### 2.3. Solar Eruptions

There is no direct evidence of a flare and solar eruption in Group #29 in Secchi's drawings. This is not surprising given that there were no systematic flare patrols in the early 1870s (Švestka & Cliver 1992) and technological limitations prevented observations of the chromosphere over the solar disk in monochromatic light. While flares were occasionally observed spectroscopically as brilliant reversals in chromospheric lines, the spectroheliograph (Hale 1892) and spectrohelioscope (Hale 1926), which permitted chromospheric imaging, had yet to be invented. A systematic flare patrol in the Hα line only started in 1934 (D'Azambuja 1934).

Nevertheless, the analyzed archival documents hint at an eruptive Sun on the days around the C-S storm given that contemporary Italian astronomers commented on enhanced solar activity observed at the solar limb with their spectroscopes on 1872 February 3–4. On February 3 Denza recorded (Denza 1872, p. 825) a significant agitation of the chromosphere and "a beautiful eruption (une belle éruption)" within "several small prominences (plusieurs petites protubérances)" at 10–11 local time (LT), viz., 9–10 UT. One of these eruptive prominences was situated "at 112°–114° from the North (à 112–114 degrés du nord)" in the southwest limb and reached "a height of 3′ (la hauteur était de près de trois minutes d'arc)" (Denza 1872, p. 825). On the same day, Secchi (1872a, p. 586) recorded a "continuation of the bright mass" at S40° E10° of the solar disk around 10 LT and described a series of prominences on the west limb between 15° and 40° south, which is consistent with Denza's observation. Secchi described the prominence activity on February 3–4, as "alive and beautiful (viva e bella)" in the margin on the right-hand side of Figure 2, like "wonderful jets (belliss. fili)," and "even if one tried, they could not make [the jets] so beautiful and even (a farlo apposta non si farebbero i peli così belli e regolari)" in his solar drawings (Osservatorio Astronomico di Roma, MS INAF 1, ff. 28–29). He also reported "a mass of high and inclined jets with holes underneath (fili alti della cromosfera e massa di fili belli ad archi intrecciati con fori sotto)" in his logbook (Secchi 1872c, p. 17). Here, the "holes" might refer to a set of prominence threads, or maybe even post-flare loops. As also cited in Secchi (1872a, p. 587), Tacchini's contemporary observation highlighted "to the west, a beautiful eruption (à l'ouest, belle éruption)" on February 3.

---

[24] This sunspot group and/or an associated coronal hole may have been recurrent for ⩾3 solar rotations, as indicated from recurrent auroral activity from early 1872 January to early 1872 March (Bhaskar et al. 2020).

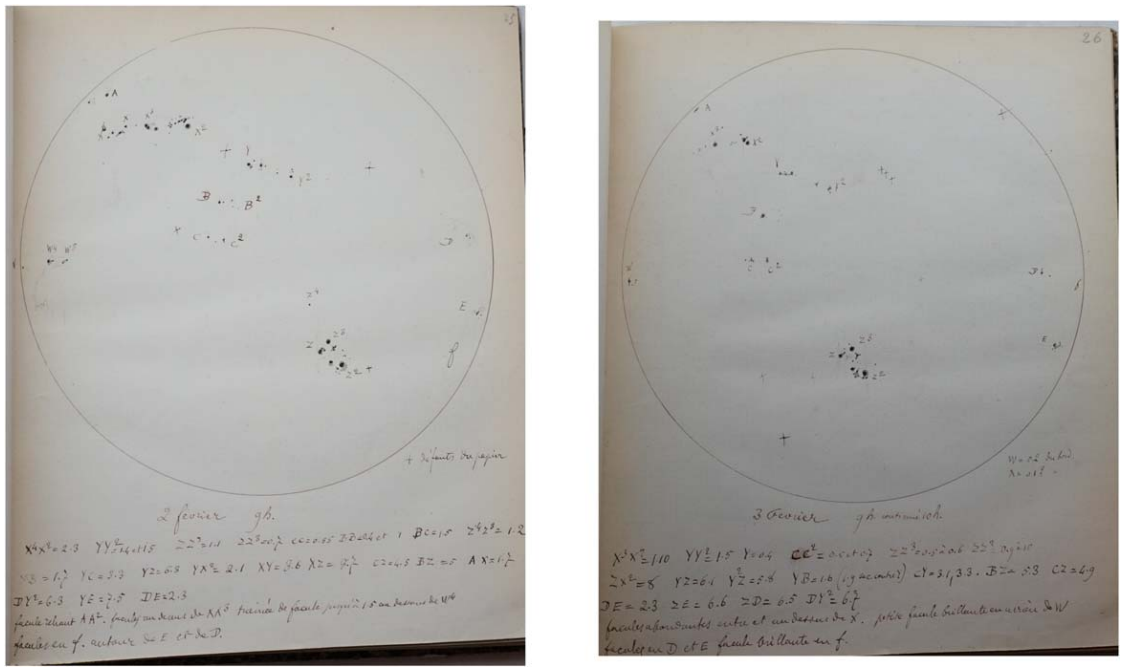

**Figure 4.** Bernaerts' sunspot drawings on 1872 February 2–3 (Source: RAS MS Bernaerts, v.3, ff. 25–26; courtesy of the Royal Astronomical Society).

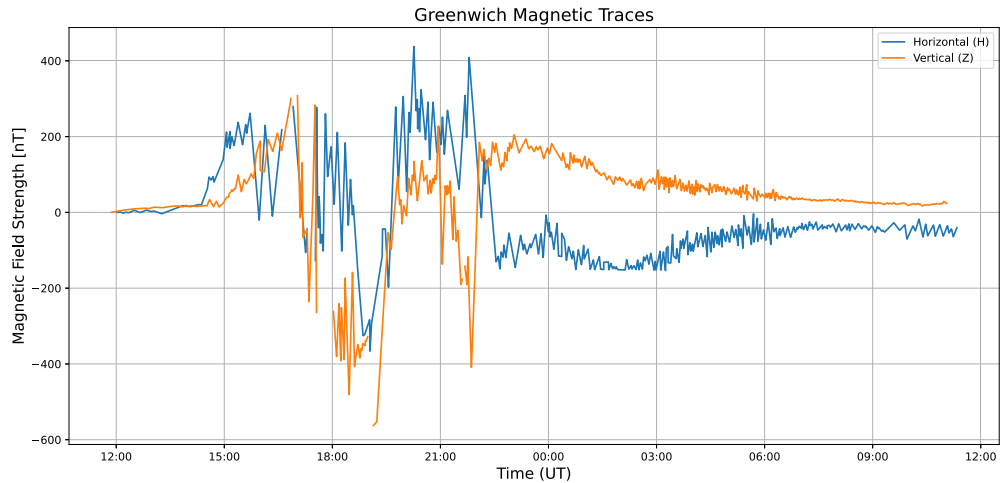

**Figure 5.** Temporal variations of magnetic field strengths at Greenwich in horizontal ($H$) and vertical ($Z$) components for the magnetic disturbances on 1872 February 4–5, which are digitised from Sheets 2 and 3 of Royal Observatory Greenwich (1872).

These reports agree reasonably well with one another on the location and timing of the eruptions observed on the southwestern limb on February 3. These eruptions may be indirectly related to the C-S storm, which began with a sudden commencement (SC) at ≈14:27 UT as shown in Figures 5 and 6 (Moos 1910, p. 452; Mayaud 1973; Silverman 2008) on February 4, ≈29 hr after the reported eruptive activity, which implies a mean propagation speed of the associated ICME of ≈1450 km s$^{-1}$. A putative identification of a source flare near disk center for the C-S storm with the prominence activity from ≈9–10 UT on February 3 is supported by both the size of the storm and the plausibly short ICME transit time of ≈29 hr for a central meridian eruptive flare.

Although the prominence observations only revealed activity at the limb, when taken together with the location of Group #29 and the timing of the onset of the great storm, this suggests the reasonable possibility that the limb activity was triggered by a major eruptive flare with an associated large-scale wave (e.g., Balasubramaniam et al. 2010). Cliver et al. (2005) reported a widespread wave (manifested by EUV dimming at the limb in an Extreme-ultraviolet Imaging Telescope (Delaboudinière et al. 1995) difference image) from a near-disk-center backside flare that was associated with a high-energy proton event in 2001 August (Cliver et al. 2005). From Figure B1 in Cliver et al. (2022b), which relates the maximum CME speed at Earth to the Sun-Earth transit time of the ICME (≈ 29 hr in this case), the likely ICME arrival speed at Earth was ≈1100 km s$^{-1}$.

### 3. Magnetic Disturbances

#### 3.1. Time Series

The storm SC indicated that the ICME-driven shock (followed by the shock-sheath structure and fast CME body) passed Earth beginning at ≈ 14:27 UT on 1872 February 4.

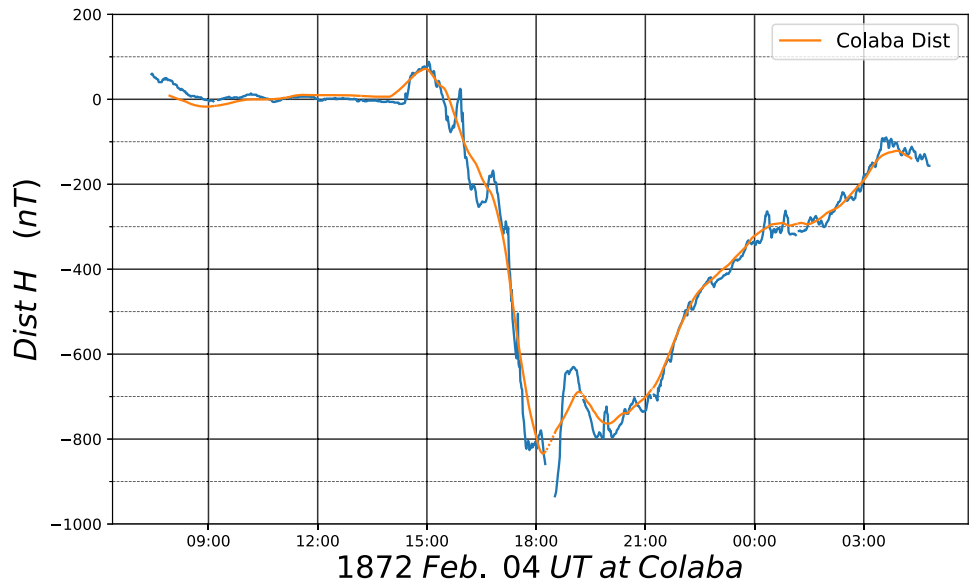

**Figure 6.** Digitization of the Colaba magnetogram on 1872 February 4–5 from Basu (1954, p. 50). For the blue curve showing the tabulated *H* data, the trace after the data gap was adjusted from that in Basu (1954, p. 50) to remove an offset that was caused by the insertion of a deflector magnet to keep the *H*-trace on scale, with the level of the trace after the insertion determined by an annotation on the original. The orange curve indicates the Dist *H* at Colaba after subtracting the Sq (solar quiet) field variation and weighting the Colaba MLAT.

The storm, recorded at the Royal Observatory Greenwich (GRW: N51°29′, E000°00′, 54°.6 MLAT; Royal Observatory Greenwich 1872; see also Valach et al. 2019),[25] was violent but of relatively short duration (≈22 hr; Jones 1955, p. 105). Figure 5 shows the digitized *H* and *Z* traces of the magnetogram from Greenwich Observatory, which was in the afternoon sector during storm onset.

Although the traces are greatly disturbed, some aspects of the storm phases can be discerned (see also Valach et al. 2019). After the shock arrival at 14:27 UT, Δ*H* shows a positive excursion lasting to just after 16 UT. This is consistent with a sudden commencement during an initial storm phase. In a statistical study, Shinbori et al. (2009) showed that the Δ*H* amplitude at St. Paratunka (46° MLAT) is 2.3 times larger than the SYM-H index at 15–16 MLT. The maximum amplitude of this positive excursion is ≈200–250 nT, on average, and the peak Δ*H* is ≈72 nT. The ratio is ≈ 2.7–3.5, which is comparable to the result of Shinbori et al. (2009). The Δ*H* amplification at mid-latitude can be explained by the contribution from the ionospheric current (Araki 1994; Shinbori et al. 2009). Subsequently there was an interval with large jerks in the Greenwich magnetogram components that lasted until ≈18:30 UT, which was followed by a negative excursion from ≈18:35 to 19:05 UT and a positive recovery (with overshoot) from ≈19:05 to ≈20:30 UT.

The oscillations between 15:00 and 16:00 UT and after 22:30 UT are notable features of the magnetogram traces in Figure 5. These low-frequency oscillations seem to be Pc-5 (Pulsation continuous-5) signals, which appear occasionally in the storm main phase and often in the recovery phase of intense geomagnetic storms (e.g., Huang 2021). Such pulsations often register in the auroral region but can also occur simultaneously at all latitudes and longitudes/local times, with a period between a few to 10 minutes. Periodic solar wind density structures are an important, but not exclusive, source of the fluctuations (Di Matteo et al. 2022). Low-latitude giant pulsations were reported in the recovery phase of the 1972 August 4–5 storm (See Figure 1(e) of Knipp et al. 2018) when solar wind speed exceeded 1100 km s$^{-1}$ (Vaisberg & Zastenker 1976). Heyns et al. (2021) used mid-latitude measurements to show that Pc-5 pulsations can drive significant Geomagnetically Induced Currents (GICs) at a mid-latitude network during intense geomagnetic storms for an extended duration. During the storm recovery phase on 2003 October 31, Pc-5 pulsations were noted in the mid-latitude Czech pipelines. During much of that day the solar wind flow speed exceeded 1000 km s$^{-1}$. Other researchers (Sakurai & Tonegawa 2005) reported the 2003 October Pc-5 pulsations as some of the largest ever recorded in the Pc-5 band. Owing to their similarity, we suggest that an 1872-level event would be disruptive to modern electric power infrastructure.

---

[25] The MLATs in this article are computed with angular distance between given observational sites and the geomagnetic North Pole on the basis of the GUFM1 model.

The negative excursion from ≈18:30 to 19:05 UT and the positive recovery with overshoot from ≈19:05 to ≈20:38 UT can also be explained by the location of the observatory relative to the ionospheric Hall current, which is known as the DP2 equivalent current (Nishida 1968). The DP2 current flows in the counterclockwise direction on the duskside and clockwise on the dawnside when one views the Northern Hemisphere from the north pole. Since Greenwich was located on the duskside, it is likely that the positive Δ*H* excursions might correspond to the equatorward part of the enhanced DP2 current flowing in the eastward direction, while the negative excursion might correspond to the poleward part of it flowing in the westward direction. The DP2 current is a manifestation of the ionospheric convection. If the convection flow reversal was located near the open-closed boundary of the magnetic field lines (Milan et al. 2017 and references therein), then it is highly plausible that Greenwich was located in the polar cap during the interval from 18:30 to 19:05 UT, as the Δ*H* was negative (see also Valach et al. 2019).

The magnetogram from the Colaba Observatory in Bombay (N18°56′, E72°50′; 10°.0 MLAT), as shown in Figure 6, was preserved as a trace copy of the original observation (Basu 1954). There is much more structure in the $H$-trace at Greenwich for the C-S storm than for that recorded at Colaba. A common feature in both magnetograms is a sharp excursion in the $H$-trace at ≈18:30 UT. At Colaba, a strong negative excursion beginning near 18:00 UT resulted in a gap from 18:17 to 18:30 UT, during which a deflector magnet was inserted to keep the $H$-trace on scale. Following this gap, the $H$-trace exhibits a positive excursion of ≈300 nT within ≈20 minutes. During the time that the $H$-trace at Colaba was recovering, that at Greenwich was decreasing by a comparable amount.

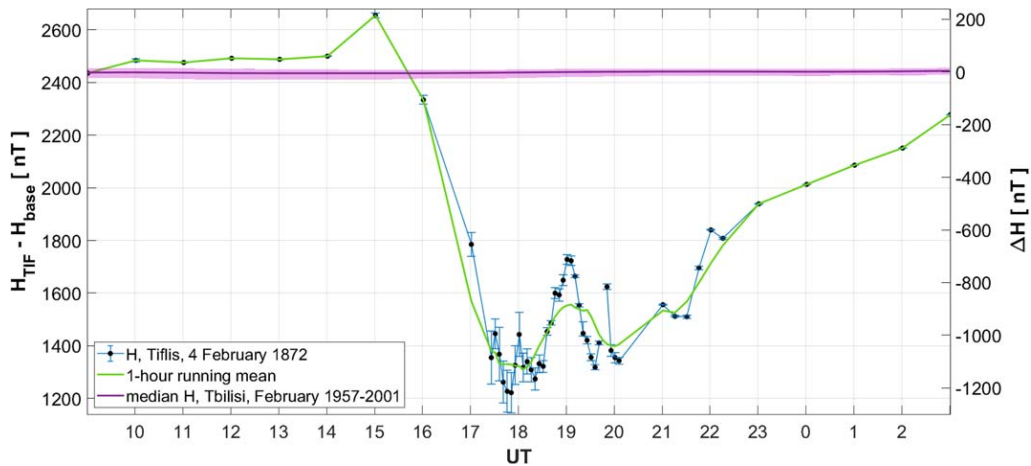

**Figure 7.** H-component measurements at Tiflis (blue curve with black circles) and one hour running mean (green curve) on 1872 February 4, as derived from Russkoe Geografičeskoe Obščestvo (1873, pp. 6–10). The purple curve with bounds indicates the diurnal curve and one sigma bounds of the H-component at Tiflis (contemporary name of Tbilisi) according to the February geomagnetic data in 1957–2001.

We have also recovered a geomagnetic measurement at Tiflis (N41°43′, E44°47′; 36.7° MLAT) from Russkoe Geografičeskoe Obščestvo (1873, pp. 6–10). This observatory recorded geomagnetic measurements in the eastward declination and in the horizontal force seemingly in a CGS unit. While the background $H$-value seems confused with that of the vertical force, the measurements' title and the actual values indicate the horizontal force. The Tiflis $H$-measurements are shown in Figure 7. These measurements are important because Tiflis is located approximately midway on a great circle line from Colaba to Greenwich. While the temporal evolution of the $H$-disturbance in the Tiflis magnetogram is similar to that in the Colaba magnetogram, the maximum amplitude in the $H$-component at Tiflis developed up to ≈1433 nT versus ≈1023 nT at Colaba (Basu 1954). The Tiflis $H$-measurements recorded the maximum at 15:01 UT (18:00 LT), a negative peak at 17:51 UT (20:50 LT), and a positive excursion up to 19:01 UT (22:00 LT).

The median diurnal curve of the $H$-component in Tiflis (thick purple curve) varies within ±5 nT according to the February geomagnetic data in 1957–2001, and can therefore be neglected in the following analysis. Quiet time $\Delta H$ apparently lies within the range of one standard deviation (purple background in Figure 7), corresponding to the values between −30 and 15 nT. For the measurements at Tiflis, the lowest geomagnetic activity with an index equal to 30 nT was observed at 9 UT. Thus, the recorded 2436 nT $H$-value at 9:01 UT can be considered to be an undisturbed field. The uncertainty of this value can be estimated as ± the width of the quiet time deviations ranges, i.e., roughly ±50 nT. We derive the positive $H$ excursion around 15 UT to be ≈219 ± 50 nT and the negative excursion around 18 UT as ≈1214 ± 50 nT, taking the local diurnal variation into account.

This storm was also recorded at Havana Observatory (N23° 08′, W082°21′, 34.°1 MLAT; Secchi 1872b). Havana sensed strong magnetic disturbances beginning at 10 LT (15:29 UT) and its magnetometer in the horizontal component went off-scale during ≈11:30–14:30 LT (16:59–19:59 UT; Secchi 1872b, p. 36). These magnetograms at Bombay and Havana (as described by Benito Viñes in Secchi 1872b) are consistent with each other in terms of temporal variation.

At Hanover VT (N43°42′, W072°17′, 55.°1 MLAT), Twining (1872, pp. 279–280) described magnetic variations in the northeastern US between 10 and 14 LT (14:49–18:49 UT): "there were three epochs of extreme and sudden deviation followed by as many of sudden change back again,—that the extreme fluctuation was 5° 40′ (in the declination) in three hours time, and that the violent disturbance preceded the visible (aurora)." The author also mentions that observations were suspended for 78 min before the local noon (15:31 UT).

### 3.2. Storm Intensity

The horizontal component, $\Delta H$, of the Bombay magnetogram in Basu (1954, p. 50) shows annotated amplitudes of the SC of 85 nT (measured at 15:03 UT from the relatively stable pre-SC trace) and an $H$ trace maximum-to-minimum (max-min) amplitude of 1023 nT (see also Moos 1910, p. 452), measured from the peak of the SC to the first measurement after the data gap from 18:17 to 18:30 UT, for which the value was chosen to give a max-min value of 1023 nT (assuming no further negative excursion during the data gap). Subtracting the SC amplitude from the maximal amplitude, the negative excursion in $\Delta H$ is estimated to be ≈−938 nT for the spot value (see Hayakawa et al. 2018). However, the annotated value of 1023 nT on which this value is based is likely to be a conservative estimate, unless the first reading from the deflected magnet was taken at the exact minimum in the storm $H$-trace.

Note that the $\Delta H$ value of ≈−938 nT is an instantaneous measurement obtained from the annotated value, whereas the Dst value should be computed by the hourly average of four mid- and low-latitude stations (Sugiura 1964; Sugiura & Kamei 1991; Akasofu & Kamide 2005; Siscoe et al. 2006; Gonzalez et al. 2011). We have further followed Dst calculation procedures to quantify the storm magnitude and time series in the Dst estimate (Dst*). These procedures require us to subtract baseline and background solar quiet (Sq) field variation from the hourly $H$ variations for four stations and average them with latitudinal weighting (Sugiura 1964). The output value is called Dist $H$. We need Dist $H$ from four mid/low-latitude stations to derive the Dst index (see Sugiura 1964). For now, we approximate the Dst estimate (Dst*) with Dist $H$, as it is extremely difficult to locate four magnetograms from mid/low-latitude regions for this storm.

Here, we have approximated the baseline with the pre-storm level and the Sq variation with the monthly $H$ diurnal variation in Moos (1910, p. 64). On this basis, we have estimated the peak Dist $H$ for the SC as $\approx 72$ nT and the minimum Dist $H$ for the storm as $\leqslant -834$ nT, respectively. Accordingly, we approximate the min Dst* of $\leqslant -834$ nT, as this value is likely to be an underestimate of the absolute value of the minimum $H$-trace excursion as noted above.

We have also computed the Dist $H$ at Tiflis. Following Figure 7, the running hourly average value around 18 UT reaches the lowest value of 1312 nT. The diurnal $H$ variation ranges $\pm 50$ nT, following sigma bounds of the H-component at Tiflis according to the February geomagnetic data in 1957–2001. Taking into account the $\Delta H$ value of $-1124$ nT (1312 nT minus the background field of 2436 nT) and the magnetic latitude of Tiflis as 36°.7, we estimate the lowest hourly Dist $H$ at Tiflis $\approx -1402 \pm 62$ nT, which is larger than the Dist $H$ at Colaba, following the Dist $H$ calculation procedure (Sugiura 1964). This can be reasonably explained by the asymmetry of the storm-time ring current (Clauer & McPherron 1980; Ohtani et al. 2007), which peaks in the dusk-midnight sector. It is not clear if the positive $H$ excursion of $\approx 219 \pm 50$ nT is entirely due to the sudden storm commencement and further investigation is needed to determine the origin of the positive $H$ excursion.

These calculations allow us to conservatively estimate the minimal Dst* for this storm as $\leqslant -834$ nT based on the Colaba magnetogram because the aurora was overhead at Tiflis and may have affected the local measurement (Russkoe Geografičeskoe Obščestvo 1873, pp. 6–10). In the following section, we reconstruct the equatorward boundary of the overhead aurora in the C-S storm down to $|24°.2|$ ILAT. From a relationship between the equatorward boundary of the overhead aurora and the minimum Dst value (Akasofu & Chapman 1963; Yokoyama et al. 1998; Hayakawa 2020; Cliver et al. 2022a) for great storms, Cliver et al. (2022a their Figure 39; see also Figure 9.4 in Hayakawa 2020) obtained a minimum Dst value of –1250 nT for 1872 storm. For now, we follow the conservative estimate for the minimum Dst* ($\leqslant -834$ nT) based on the Colaba magnetogram for the C-S storm.

## 4. Aurorae

### 4.1. Auroral Visibility (MLAT) and Equatorward Boundary of the Auroral Oval (ILAT)

This great magnetic disturbance was accompanied by significant auroral displays down to extremely low magnetic latitudes. We have investigated reports for auroral observations on 1872 February 4–6, following three contemporary auroral compilations for this storm (Preece 1872; Fron 1872; Donati 1874); auroral compilations from newspapers, diaries, and chronicles in South Asia and East Asia (Chapman 1957b;

Hayakawa et al. 2018); the Yearbooks of the Russian Central Observatory (Wild 1874) and the American Chief Signal Office (ARSCO 1873); contemporary major scientific magazines [Nature (Nature 1872a, 1872b; Nature 1873), Meteorological Magazine (MMA 1872), Quarterly Journal of the Royal Meteorological Society (Royal Meteorological Society, 1873), American Journal of Science (Dana & Silliman 1872a, 1872b), Comptes Rendus (Académie des Sciences 1872), Wochenschrift zür Astronomie, Geographie, und Meteorologie (Heis 1873, 1874, 1876), and Zeitschrift der Österreichische Gesselschaft der Meteorologie (Jerinek & Hann 1872)]; newspapers in India, Australia, Iberian Peninsula, Azores, Bermuda, and the USA; several ship logs; and other miscellaneous reports (Tristram 1873; Capron 1879; Rixo 1994, p. 510; Al-Kurdī al-Makkī 2000, p. 370).

Figure 8 shows the distribution of auroral observation sites on 1872 February 4–6 based on the contemporary reports of visible aurorae. Aurorae were credibly reported for several locations below $\approx 20°$ MLAT, extending down to magnetic latitudes such as Bombay (10°.0 MLAT) and Khartoum (13°.4 MLAT), which are lower than the equatorward extent of visibility reported for either the Carrington storm in 1859 ($|17°.3|$; Hayakawa et al. 2020) or the 1921 May storm ($|16°.2|$; Angenheister & Westland 1921; Silverman & Cliver 2001). We can reconstruct the equatorward boundary of the auroral oval during the Chapman–Silverman storm based on the elevation angle of the auroral display and the magnetic latitude of the observational sites by assuming that the auroral emissions extended up to 400 km along the magnetic field lines (Roach et al. 1960; Silverman 1998; Ebihara et al. 2017). Accordingly, for events with elevation-angle reports, we can compute the lowest ILAT of overhead aurora, which is identical along the magnetic field line from the magnetic footprint (O'Brien et al. 1962), as shown in Figure 9, rather than the lowest MLAT of auroral visibility, which is the lowest latitude at which the aurora was actually observed. We selected the seven most credible low-latitude records for which an ILAT could be estimated (based on the detail/precision of the auroral reports) and summarize their details in Table 1. From these auroral records, the equatorward boundary of the auroral oval is considered to extend down to 24°–32° ILAT across the East Asian, Indian, African/European and eastern American sectors, i.e., $\approx$E140° to $\approx$W060°.

In the East Asian sector, the auroral displays were widely reported in Japan, Korea, and China; as exemplified in Figure 10 (Hayakawa et al. 2018). Among them, Mr. Vignale, General Consul of Italy in Shanghai, reported that "the bright arc extended almost for 50° very close to the zenith" at Shanghai (Donati 1874, p. 8). Considering the magnetic coordinates (N31°14′, E121°29′, 19°.9 MLAT), we obtain an ILAT of $\approx$ 24°.2 ILAT for this location (Hayakawa et al. 2018).

In the Indian subcontinent, auroral emissions were reported in conjunction with telegraph disturbances between Jacobabad (N28°17′, E068°26′, 19°.9 MLAT) and Lahore (N31°35′, E074°19′, 22°.4 MLAT). At Jacobabad, the auroral display was "shooting from the east horizon to the zenith and very nearly at right angles to the magnetic meridian" and "after an hour there appeared in the aurora arch a little below the zenith a bright blue light of a dome shape quite intense; after some 15 minutes it suddenly dissolved to a deep violet" (Times of India, 1872-02-15, p. 2; Chapman 1957b, p. 187). Therefore, in this

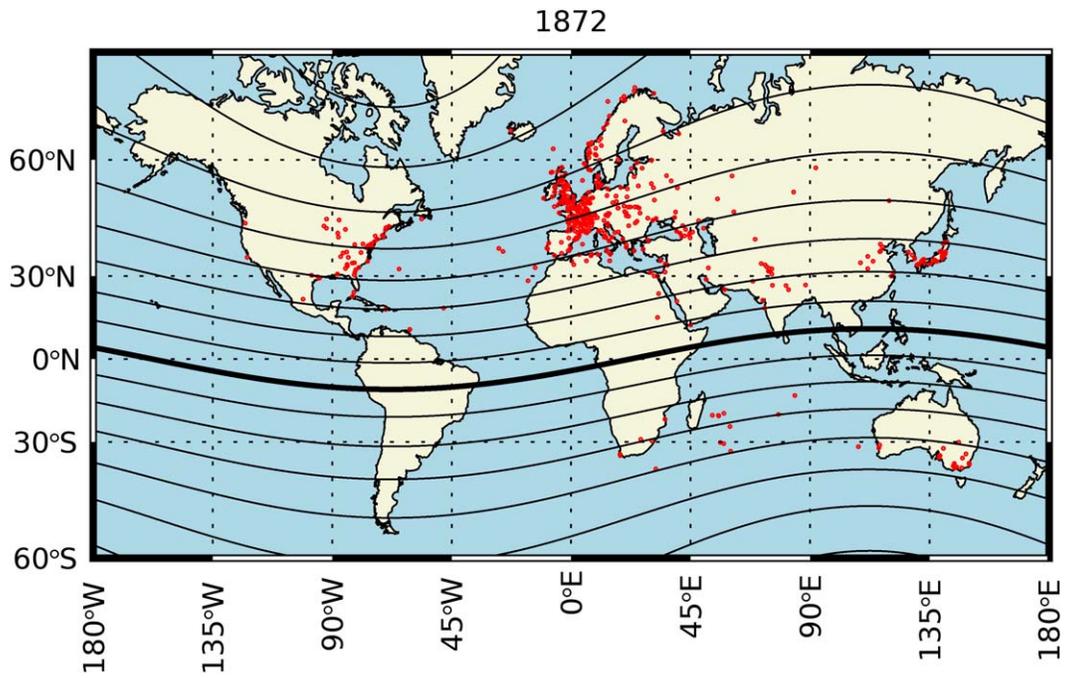

**Figure 8.** Auroral visibility on 1872 February 4–6. Each red dot indicates a location for which auroral visibility was reported. The contour lines indicate the magnetic latitude (MLAT) with an interval of 10°, which are based on the magnetic poles for 1872 determined by the model GUFM1 (Jackson et al. 2000).

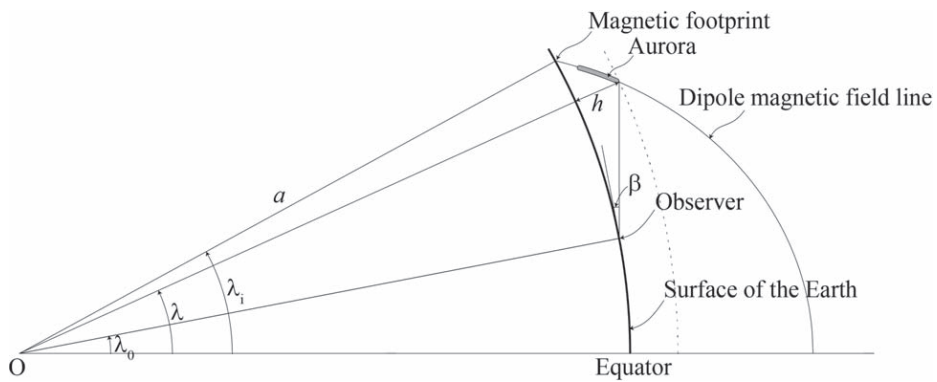

**Figure 9.** MLAT of auroral visibility $\lambda_0$, MLAT of the topside of aurora $\lambda$ and ILAT of the aurora $\lambda_i$ (reproduced from Figure 2 of Hayakawa et al. 2018 with permission).

**Table 1**
Locations for the Most-equatorward Credible Observations of Aurora for the 1872 Storm from which an ILAT Could Be Determined

| Site | Latitude | Longitude | MLAT | Max. in UT[a] | Elevation | ILAT |
|---|---|---|---|---|---|---|
| Shanghai | N31°14′ | E121°29′ | 19°9 | 17:54 | Zenith | 24°2 |
| Jacobabad | N28°17′ | E068°26′ | 19°9 | 18:56 | Zenith | 24°2 |
| Mauritius | S20°10′ | E057°31′ | −26°3 | 19:25 | 72° | 30°5 |
| St. Denis | S20°52′ | E053°10′ | −26°3 | 18:27–19:27 | Zenith | 29°6 |
| Ispahan | N32°39′ | E051°40′ | 26°6 | N/A | 60° | 31°6 |
| Sifook | S22°16′ | E035°21′ | −24°3 | N/A | Zenith | 27°9 |
| Courland Bay | N11°13′ | W060°47′ | 22°6 | N/A | Halfway | 29°2 |

**Note.**
[a] This time can be either for sudden onset of the aurora, rapid change, or maximum sky coverage (i.e., climax of the event).

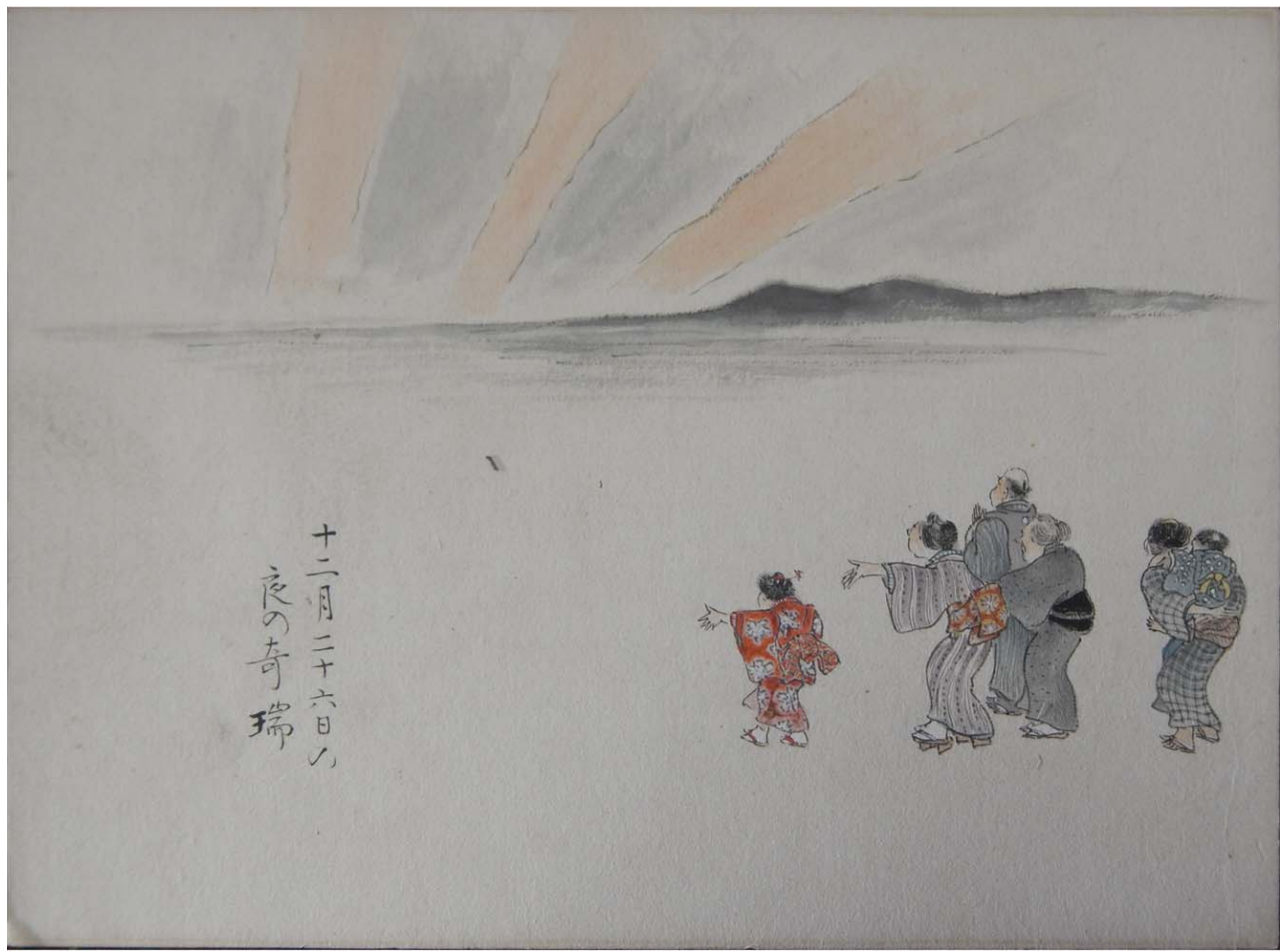

**Figure 10.** A Japanese drawing showing an auroral display seen at Okazaki (N37°54, E137°10, 24°4 MLAT) for the 1872 February 4 storm, which is preserved in the Shounji Temple, as reproduced from Hayakawa et al. (2018), with permission of the Shounji Temple.

sector, we regard the reported lights not as stable auroral red (SAR) arcs (see Kozyra et al. 1997) with a monochromatic reddish appearance but as auroral displays themselves that extended down to 24°2 ILAT, and which were visible at Jacobabad (19°9 MLAT) over the zenith.

Heading westward, several low-latitude aurorae were reported around the Indian Ocean. In the southern hemisphere, aurorae were reported up to 72° in elevation at Mauritius (S20°10′, E057°31′, −26°3 MLAT) and up to zenith at St. Denis (CR65: S20°52′ E053°10′, −26°3 MLAT). On this basis, the auroral oval was reconstructed down to 30°5 ILAT and 29°6 ILAT, respectively. At St. Denis, "the colored part [of the aurora] to the south rose to the zenith" and "immense luminous jets [rose] from the horizon to the zenith, like columns of fire, bands, or flares of a whiter light" with significant brightness (Vinson 1872, p. 721). Here also, the aurora does not fit the description of an SAR arc. In the Middle East, several aurorae were reported in Iran (Preece 1872). At Ispahan (MI17: current Esfahān, N32°39′, E051°40′, 26°6 MLAT), the "whole of the Northern sky was crimson up to about 60°, like a magnificent sunset" (Preece 1872, p. 111). This indicates the equatorward extension of the auroral oval down to 31°6 ILAT, whereas this might alternatively indicate an equatorward boundary of the SAR arc based on the reported colouration and morphology.

On the eastern coast of Africa, the aurora was reported down to Sifook (S22°16′, E035°21′; −24°3 MLAT) in current Mozambique, where the aurora was "almost extended to the zenith itself at one time in a broad belt" (Erskine 1875, p. 91). Considering the MLAT of Sifook and the reported elevation angle, we consider the auroral oval to have extended down to 27°9 ILAT.

In the American sector, the aurora was reported down to Tobago: "Mr. Taylor, master of the barque Tobago was riding at anchor at Courland Bay, and was a witness of this aurora. He described it … as of 'a dark-red colour, extending half way up to the zenith and very brilliant, its situation being about NW by W." We interpret this report as an auroral display or an SAR arc extending 45° in elevation at Courland Bay of Tobago (N11°13′, W060°47′, 22°6 MLAT; Yeates 1872) off the northern coast of Venezuela; the auroral emission region is considered to have extended down to 29°2 ILAT.

Here, the reports at Esfāhan and Courland Bay describe monochromatically reddish glows. These glows may be SAR arcs given that only the red color is reported. SAR arcs have a dominant red color and are formless features of long duration (Cornwall et al. 1970, 1971; Kozyra et al. 1997). However, Jacobabad witnessed both the dynamics ("immense luminous jets [rose] from the horizon to the zenith") and coloration (as "a bright blue light of a dome shape") as reported in the Times of India (1872-02-15, p. 2; see also Chapman 1957b, p. 187) of the C-S aurora reported from other locations defy explanation in terms of SAR arcs. The aurorae at Sifook, Mauritius and St.

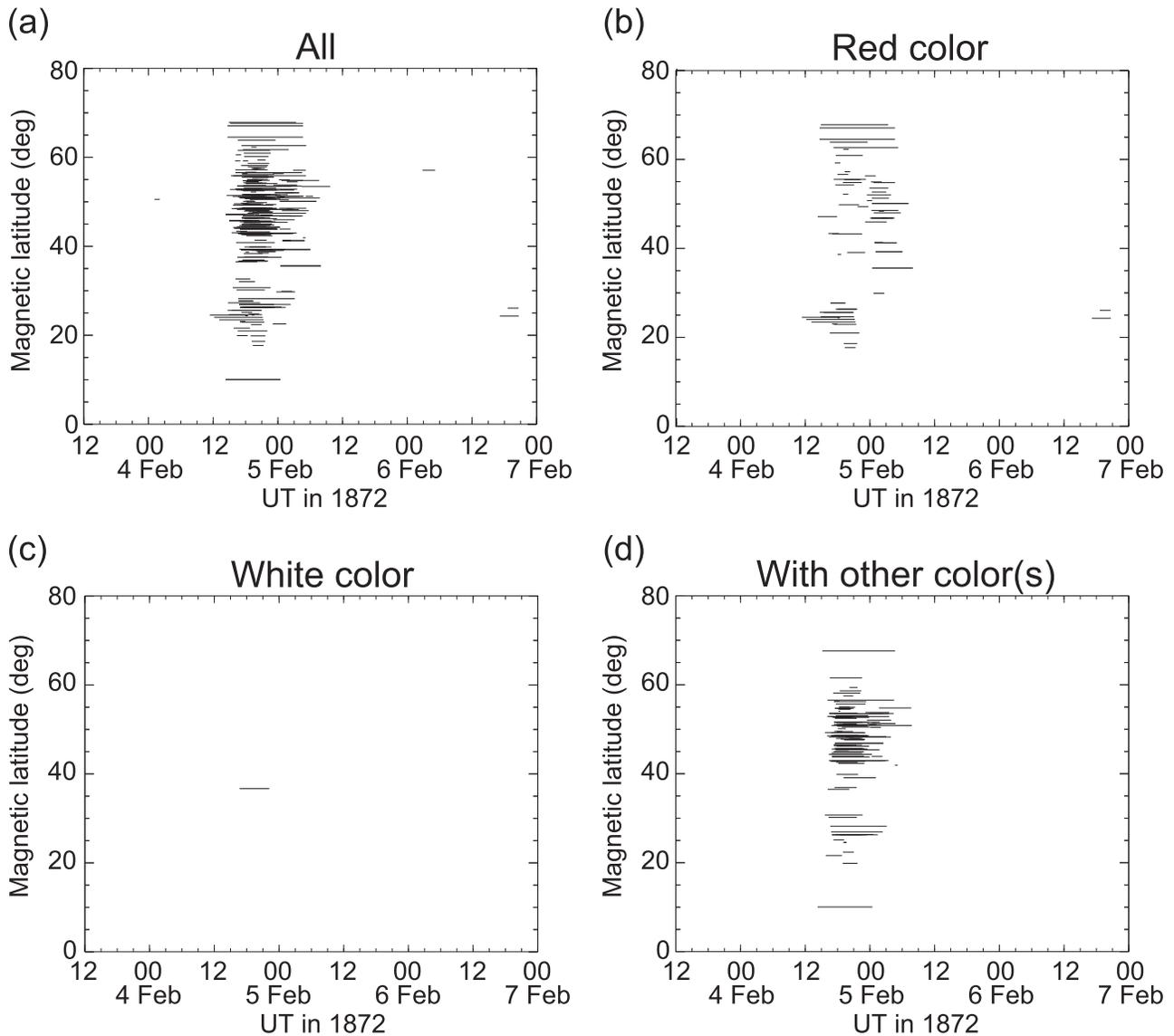

**Figure 11.** Duration of the auroral visibility against the absolute value of magnetic latitude. Each panel shows visibility reports (a) in all, (b) only in reddish color, (c) only in whitish color, and (d) with non-reddish colors too (occasionally including reddish color as well). Some reports do not have color descriptions. They appear only in panel (a).

Denis were also reported with whitish and purplish structures (Erskine 1875, p. 91; N25: Meldrum 1872a, p. 392; Vinson 1872, p. 721). Therefore, we consider that the auroral oval extended down to 24°–30° ILAT in its maximal intensity in each geographical sector.

### 4.2. Temporal Evolution of Aurora and Comparison with Magnetograms

Figure 11 shows the auroral visibility on 1872 February 4–5 as a function of magnetic latitude and time, with sites of auroral visibility ranging from 10° to 68°. The aurora was first noted by several observers near the SC time of ≈14:27 UT (as was the case for the 1989 March storm; Boteler 2019) or even earlier. For three stations (one in Korea and two in Japan), the reported auroral onset was ≈12 UT. Given the lack of geomagnetic activity preceding the SC, we suspect that these time stamps were erroneous because of lack of precise timing in East Asian records from the regular citizens during this period.

The bulk of the plotted observers witnessed aurora at ≈20 UT and observations from Europe mostly ended by 24 UT, with only a few stations reporting aurorae after 6 UT on February 5. Note that the equatorward boundary of auroral visibility is not necessarily the same as that of the auroral oval because the aurora can be seen above the horizon, looking in most cases (see below) in the direction of the magnetic pole, given the auroral altitudes of ≈400 km. In particular, Figure 11(d) shows auroral visibility with non-reddish color around the negative Dist $H$ peaks at ≈18 UT in Colaba and Tiflis. This concentration is interesting because these reports indicate that they are not SAR arcs but the robust aurorae on the basis of the inclusion of non-reddish colorations (see, e.g., Kozyra et al. 1997).

Here we consider the timings of dramatic onset of the aurora (e.g., Jacobabad), rapid change (Barnstaple), or maximum sky coverage (auroral climax) for comparison with magnetic records from Havana, Greenwich, Tiflis, and Colaba, with a focus on the low-latitude auroral observations from Table 1.

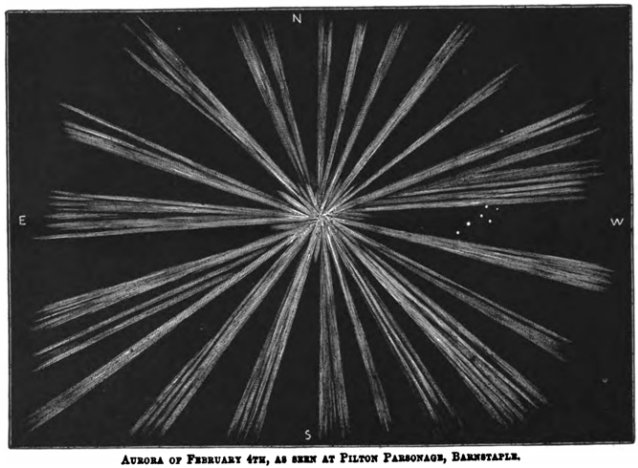 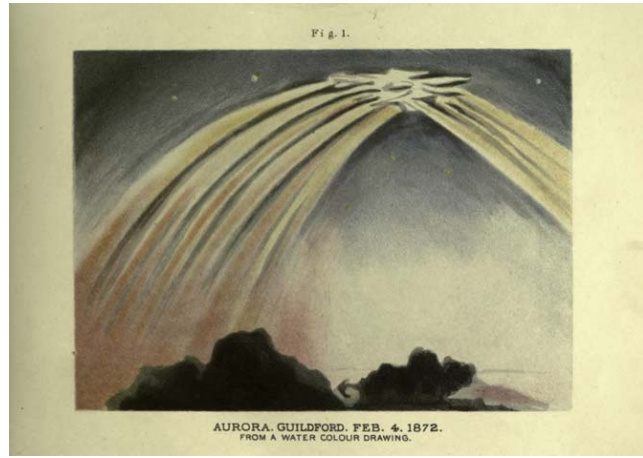

**Figure 12.** Corona aurorae witnessed in Barnstaple (Hall 1872, Frontispiece) and Guildford (Capron 1879, Plate IV) on 1872 February 4.

Observers in East Asia witnessed aurorae during the earliest phase of this storm. The majority reported auroral onset around local midnight (14–15 UT) or a little earlier, with a ±1 h uncertainty in their local time system (Uchida 1992). The description of the aurora from Shanghai is from Von Gumpach:[26] "The aurora was observed in Shanghai as a far-away explosion, of rather weak intensity. The sky at the horizon appeared dark, and the glowing arc covered almost 50°, near the zenith. We do not have an exact time for the beginning of the phenomenon. Its maximum was reached at around 2 am [17:54 UT] on February 5. The aurora stopped at 2:30 [18:24 UT], or shortly thereafter" (Donati 1874, p. 8).

In the Indian Subcontinent, the following account of the aurora from Jacobabad is from a correspondent to the Times of India (1872-02-15, p. 2): "As I was returning home about half-past 11 p.m. [18:56 UT], a sudden change from darkness to light was noticed as bright as the full moon. I was amazed; the conversion of Saint Paul came vividly before me—in fact, I was terrified by the sudden-change; my dog became motionless and seemed, to tremble. I thought it must have been a fire, no, the whole place was magically illuminated. … After an hour there appeared in the Aurora arch a little below the zenith a bright blue light of a dome shape quite intense; after some 15 minutes it suddenly dissolved to a deep violet. The sky was clear and the phenomena was observed very distinct; it traveled towards the south and lasted till 2–15 (P.M.) [mistaken from A.M.: 21:41 UT] this morning, a duration of nearly 2 3/4 hours." The sudden onset of the aurora at 18:56 UT in Jacobabad occurred near the ≈19:01 UT end of the recovery (with overshoot) of a lower-limit (because of a data gap) > 150 nT negative excursion in the $H$ trace at Colaba (Figure 6) that necessitated the insertion of the deflector magnet during the data gap from 18:17 to 18:30 UT, and near the 19:03 UT minimum of the Greenwich $H$-trace and onset of the data gap in the corresponding $Z$ magnetogram (Figure 5). The $H$-trace behavior at Colaba (Figure 6) was mimicked by that in the Tiflis magnetogram (Figure 7), which recorded a negative $H$ peak at 17:51 UT, and subsequent positive excursions from 17:51 to 18:01 UT of 221 nT and from 18:21 to 19:01 UT of 449 nT.

Observers in the southern Indian Ocean witnessed a great extension of the aurora australis. From the ship Pendragon (S13°43′, E084°13′): "At midnight [≈18:23 UT] very suspicious-looking weather to the S., the sky being quite red" (Meldrum 1872b, p. 29).

The aurora as observed from Mauritius (S20°10′, E057°31′) by Meldrum (1872b): "9.48 p.m. [17:58 UT]—An irregular convex arch of dark red light extending over about 60° of the horizon, and having its vertex in the line of the magnetic meridian. … 11 p.m. [19:10 UT]—Sixteen luminous bands of a steel gray to a silver white color, extending from as low down as I can see to within 20° of the zenith. … 11.6 p.m. [19:16 UT]—The parallel bands are still seen. They cover the greater part of the hemisphere, extending (at the meridian) to about 72° above the horizon. … 11.15 P.M. [19:25 UT]—A deep red glow from E. to W. by S. along the horizon. Fourteen parallel bands of a silvery color, with dark bands between them. They lie south and north, occupying nearly the whole southern hemisphere as far as the eye can reach, and are flanked at east and west by patches of blood and cherry red. … 1.20 A. M. [21:30]—A bright red glow from S.E. to S.W. Intensest [sic] below the Centaur. Soon died away."

The aurora as observed at St. Denis (Réunion; S20°53′, E055°27′) by Vaillant (1872): "This was a magnificent aurora australis like I had never seen before. The first glows appeared around 8:30 [16:57 UT] … its greatest intensity was reached between 11:30 in the evening [19:57 UT] and 1:30 in the morning [21:57 UT]." From the report by Vinson (1872): "This beautiful aerial phenomenon started around 8:30 in the evening [16:57 UT] … From 10 to 11 [18:27–19:27 UT], the aurora seemed to reach the maximum of its intensity."

In the European sector, the auroral display was already visible after ≈16–17 UT (Earwaker 1872) from the southern horizon, and expanded to the northern sky with developments of distinct corona and rays (Figure 12; Earwaker 1872, pp. 322–324; Capron 1879, pp. 19–21). Townshend M. Hall (1872), a fellow of the Royal Geological Society, reported a "southern aurora" as "6. 0 p.m. [18:16 UT] Diffused red light near Orion's belt, in S.E. 6.15 [6:31 UT]" and "Several white rays in the South, diverging from near the Pleiades" at Barnstaple (N51°05′, W004°04′, 55.°0 MLAT): "[up] to 6.55 [19:11 UT], no auroral light was distinguishable in the northern portion of the sky; but at that moment there was a sudden outburst of rays from the central point, covering the entire heaven in every quarter, several of the rays in the E. and E.N.E. being, however, especially remarkable for their width and color. The sky remained absolutely without a cloud until nearly 8 o'clock [20:16 UT], when there were only the remains of a few white streaks visible, and an hour later rain clouds, with a thick mist, obscured even the brightest of the stars" (Hall 1872, p. 2). The rapid expansion of the aurora to the northern sky at 19:11 UT at Barnstaple occurred within minutes of the lowest excursions in the Greenwich $H$- and $Z$-traces and near the peak of the positive bay following the 18:30–18:47 UT data gap in the Colaba $H$-trace, and falls near the sudden onset of the bright auroral display at Jacobabad at 18:56 UT.

---

[26] Von Gumpach was a professor of mathematics and astronomy at Peking (Beijing) College from 1866 to 1868, and then moved to Shangai until his death in 1875 (Donati 1874, p. 8; Le Conte 2019, pp. 9–10).

The aurorae were visible in the American sector only after the end of the storm's main phase. US weather maps for the local afternoon and evening of 1872 February 4 (16:35 LT and 23:35 LT in Washington, DC) show that the reporting stations for the US mountain states were under mostly cloudy skies, with snow obscuring the sky in several locations.[27] A winter storm covered most of the inter-mountain west. The 23:35 LT synoptic (Washington, DC) discussion notes that aurora had been reported in New York state and at the Mobile, Alabama station. A newspaper report also noted a "splendid" overhead aurora observed in Austin, Texas (The Democratic Statesman, 1872-02-06, p. 3). In sum, aurorae were reported in the US, weather permitting, and the lack of auroral reports in much of the western US (See Figure 8) was likely to have been weather-related. Auroral visibility was reported down to the Courland Bay in Tobago (N11°11′, W060°44′, 22°.5 MLAT), where it was witnessed at 19:00–21:30 LT (23:03–25:33 UT; Yeates 1872). This is consistent with a low-latitude auroral observation during the recovery phase of the storm, just prior to local sunset at Courland Bay (sunset ≈ 22:05 UT).

These comparisons of temporal auroral evolution at low magnetic latitudes and recorded magnetic activity indicate a close timing correspondence between key features in the evolution of both phenomena for the 1872 aurora. In particular, we note that the sudden appearance at 18:56 UT of aurora at Jacobabad (MLAT=19°.9) occurred during the $H$-trace recovery from 18:31 UT—following a 14 minute gap during which a deflector magnetic was inserted at Colaba (10°.0 MLAT) to keep the $H$-trace on scale during a sharp negative excursion—to a local peak at ≈19:01 UT. Corresponding behavior was observed in England (approximately five hours in local time and 40° distant in MLAT) where the aurora as viewed from Barnstaple (MLAT=55°.0 N) suddenly filled the entire sky at 19:11 UT, near the minimum in the Greenwich (MLAT=55°.0 N) $H$-trace at 19:03 UT. This is consistent with the Greenwich magnetogram that plausibly indicated Greenwich being located in the polar cap during the interval from 18:30 to 19:05 UT (Section 3.1).

## 5. Reports of Extremely Low Magnetic Latitude in 1872 February

The 1872 February aurora is notable for the number of reports of aurora from low magnetic-latitude (<|20|° MLAT) locations. Table 2 lists 13 such sites. In contrast, only one such sighting is documented for the both the 1859 September (vessel Dart; |17°.3| MLAT) and 1921 May (Apia; |16°.2| MLAT) storms.

---

[27] https://library.oarcloud.noaa.gov/docs.lib/htdocs/rescue/dwm/1872/18720204.pdf

### 5.1. The Aurora at Bombay

In the 1872 February 6 issue of the Times of India, the aurora at Bombay was described as follows: "After sunset … the aurora was slightly visible … [and] It was distinctly visible until sunrise."

This report of a long-lived aurora at 10°.0 MLAT was viewed with skepticism by Silverman (1995, 2006), although, as noted above, Silverman (2008) subsequently reckoned that the report from Bombay was credible. Silverman's initial difficulty in accepting the report of aurora in Bombay was based on the likely conflation of telegraph outages with a reported aurora at Singapore (−10°.0 MLAT) for the great storm in 1909 September (Silverman 1995; Hayakawa et al. 2019a). The ≈10° gap in corroborating reports from sites between Bombay and Jacobabad may have been an additional factor. Our current investigation, however, indicates that is no longer the case for the Bombay report. Table 2 includes five stations between Bombay and Jacobabad on the ≈ W075° meridian, for which aurorae were reported with corroborating details. The MLATs for these five stations are well distributed between 12°.9 MLAT and 17°.6 MLAT (Table 2). While the timing of the aurora from sunrise to sunset at Bombay is problematic given the discordant detailed report from Jacobabad, the likely source of the report of aurora at Bombay (and also Aden; see Section 5.2) in the Times of India article appears to be beyond reproach.

George Benham Stacey (1839–1904) was one of the few people in Bombay on 1872 February 4 who would have had both the knowledge and motivation to look for an aurora in conjunction with the onset of the main phase of the great magnetic storm. He had a remarkable ≈ 50 yr career in the telegraph service, which began at age 12 when he joined the Electric Telegraph Co. His obituary (Anonymous 1904) details his rapid ascent up the ranks of his profession with postings at the Hague, Berlin, London, and Egypt, before a final appointment as the manager of the Eastern Telegraph Co. station in Bombay in 1872, where he remained for 28 yr. Quoting from his obituary, "Mr. Stacey was an ideal representative of the great cable service … and no man in the service has ever been more universally respected alike by the staff and the directors of the companies in which he was engaged." In Preece (1872), Stacey's report from Bombay is succinct: "Strong earth currents 7.30 p.m. on 4th February. Continued until 7.0 a.m. on 5th. Aurora plainly visible 8:30 p.m. until 4:30 a.m." This report reduces some of the timing discrepancies in the auroral reports between Bombay and Jacobabad and the "plainly visible" repeats the text in the opening sentence of the February 6 Times of India article that appears to be an amalgamation from multiple sources (see Appendix). This report is also significant, reporting "strong earth currents" in Bombay (10.0° MLAT) immediately after the storm peak.

In sum, the evidence of corroborating reports between Bombay and Jacobabad, and the tribute to Stacey, combined with triangulation analyses by Chapman (1957b) and Hayakawa et al. (2018) that indicated that the overhead aurora inferred for Jacobabad could have been observed from Bombay at an elevation angle of 10°–15°, and new evidence for additional low-latitude aurorae visible at other longitudes as discussed in the following section, tip the scales in favor of the reality of the observation of aurora from Bombay in the C-S event.

Table 2
Geographical Distribution of the Reported Auroral Visibility in the Most-equatorward Sites (<|20.0|° MLAT), Shown from the East to the West According to Their Longitudes

| Site | Latitude | Longitude | MLAT | Reference | Volume | Page |
|---|---|---|---|---|---|---|
| Shanghai | N31°14′ | E121°29′ | 19°.9 | Donati (1874) | 1874 | 8 |
| Shàoxīng | N30°00′ | E120°35′ | 18°.7 | Hayakawa et al. (2018) | 862 | 9 |
| Darjeeling | N27°02′ | E088°16′ | 16°.6 | The Englishman's Overland Mail | 1872-02-16 | 3 |
| Allahabad | N25°26′ | E081°52′ | 15°.5 | The Friend of India | 1872-02-15 | 16-17 |
| Lucknow | N26°50′ | E080°55′ | 17°.0 | The Englishman's Overland Mail | 1872-02-16 | 3 |
| Jeypore | N26°56′ | E075°49′ | 17°.6 | The Friend of India | 1872-04-25 | 14 |
| Bombay | N18°56′ | E072°50′ | 10°.0 | The Times of India | 1872-02-06 | 2 |
| Bombay | N18°56′ | E072°50′ | 10°.0 | The Homeland Mail | 1872-03-04 | 217 |
| Bombay | N18°56′ | E072°50′ | 10°.0 | The Englishman | 1872-02-10 | 2 |
| Bombay | N18°56′ | E072°50′ | 10°.0 | Donati (1874) | 1874 | 8-9 |
| Bombay | N18°56′ | E072°50′ | 10°.0 | Fron (1872) | 1872-03-19 | 3 |
| Bombay | N18°56′ | E072°50′ | 10°.0 | Nature (1872a) | 5 | 323 |
| Bombay | N18°56′ | E072°50′ | 10°.0 | AJS | 103 | 391 |
| Bombay | N18°56′ | E072°50′ | 10°.0 | HEIS | 15 | 119 |
| Bombay | N18°56′ | E072°50′ | 10°.0 | OGM | 7 | 134 |
| Bhownuggur | N21°46′ | E072°09′ | 12°.9 | The Times of India | 1872-02-10 | 3 |
| Bandar-e-Jask | N25°29′ | E057°47′ | 18°.6 | Fahie (1872) | 13 | 254 |
| Jacobabad | N28°17′ | E068°26′ | 19°.9 | The Times of India | 1872-02-15 | 2 |
| Aden | N12°49′ | E045°02′ | 8°.3 | The Times of India | 1872-02-06 | 2 |
| Aden | N12°49′ | E045°02′ | 8°.3 | The Homeland Mail | 1872-03-04 | 217 |
| Mecca | N21°25′ | E039°49′ | 17°.7 | Al-Kurdī al-Makkī (2000) | 4 | 370 |
| Khartoum | N15°35′ | E032°32′ | 13°.4 | Pictet (1872) | 1872 | 74 |

### 5.2. Reported Aurorae at Aden, Khartoum, and Gondokoro

The Times of India article from 1872 February 6 on the aurora ends with the sentence, "At Aden [8°.3 MLAT] the aurora was brilliant in the extreme." Fron (1872) writes, "[The Aurora] was seen in Cairo, Egypt [27°.8 MLAT], in Chartum [Khartoum; 13°.4 MLAT], in Upper Egypt and probably—according to Mr. Raoul Pictet[28]—up to Gondokoro[29] [3°.1 MLAT]". Fron (1872) annotated the tabulated Gondokoro observation with a question mark (?). Silverman (2008) could not uncover Pictet's original report and accordingly doubted the reliability of the records of aurora at these three stations, owing to the lack of specific details.

Our archival investigations have successfully located Pictet's original report in *Archives des Sciences Physiques et Naturelles* (Pictet 1872), where he reported telegraph disturbances and enhanced geoelectricity between Cairo and Khartoum in the local afternoon on 1872 February 4, Khartoum's inquiry to Cairo about the witness of a great reddish light on the horizon in that night, and his intention to search for further possible auroral visibility in Gondokoro. His report translates as follows: "I am coming back from the telegraph (office) where I collected the following information about the electrical phenomena that occurred yesterday evening. Yesterday in the afternoon, it was difficult to communicate with Khartum, which is located 15° North, 35° of East longitude. The [communication] devices were 'talking' by themselves: signs that were sent out were missing. The ground currents strongly impeded the service and the employees were fully confused, as they could not figure out what was causing those disorders. Yesterday evening, the office in Cairo got a dispatch asking what was the big red glow that was seen at the horizon, and suspecting a big fire. The telegraph line does not continue further South, beyond Khartum, but it is probable that this aurora has also be seen up to Gondokoro, at 5° North latitude. I shall inquire about it and I will confirm to you if they did see the aurora" (Pictet 1872, p. 74).

According to this report, the telegraph disturbance between Khartoum and Cairo is in at least rough temporal alignment with the climax of the geomagnetic storm in Figures 5–7. The auroral report from Khartoum looks realistic given that Khartoum confirmed the witness of great reddish light and inquired about its nature to Cairo. In contrast, Pictet only speculated a possibility of further auroral visibility down to Gondokoro without confirmation. This is probably why Fron assigned a question mark to Gondokoro in his list. Without independent confirmation, this is no more than Pictet's speculation. From Gondokoro, the auroral display could have been visible only up to 3° in altitude by triangulation, assuming the equatoward boundary of the auroral oval as ≈ 24°.2 ILAT in this meridian as well.

---

[28] Raoul-Pierre Pictet (1846–1929) was a physicist from Switzerland, who (independently, along with Louis-Paul Cailletet) produced liquid oxygen in 1877. Pictet was an instructor at Ecole Superieure in Cairo from ≈1871 to 1875 (O'Conor Sloane 1920, p. 154).
[29] This town is now known as Juba in South Sudan.

On this basis, we confirm the auroral visibility down to Bombay (10°.0 MLAT) and favor auroral witness down to Khartoum (13°.4 MLAT). The brief report regarding Aden (8°.3 MLAT) lacks details but was attributed to Stacey in Preece (1872). It is reasonable to assume that Preece was in communication with a local witness at Aden. Like the report of aurora at Singapore in 1909, the Aden report could be based on conflation with the disruption of telegraph communications because it was the first station after Bombay on the east–west portion of the link to London (Supplement to the Electrician 1894, p. 11). Based on triangulation, the aurora would be visible at these two sites (Aden and Khartoum) according to the equatoward boundary of the auroral oval. If we expect the equatoward boundary of the auroral oval to be ≈24°.2 ILAT based on the reports from Shanghai and Jacobabad, then the calculated elevation angles at Aden and Khartoum would be 10°.9 and 25°.0, respectively. A report from Mecca (N21°25′, E39°49′, 17°.7 MLAT; Al-Kurdī al-Makkī 2000, p. 370; see also Bekli & Chadou 2020) reduces a large gap in magnetic latitude between Aden (N12°49′, E45°02′, 8°.3 MLAT) and other observational sites in West Asia such as Syene (N24°05′, E32°56′, 21°.6 MLAT), Suez (N29°57′, E32°34′, 27°.4 MLAT) and Sebbeh (Tristram 1873: N31°19′, E35°22′, 28°.2 MLAT). These details make the Aden report credible, in our opinion.

## 6. Three Extreme Geomagnetic Storms: 1859 September, 1872 February, 1921 May

The minimum storm intensity for the Chapman–Silverman storm is conservatively estimated to be minimum Dst ⩽ −834 nT (running hourly average; single station). This is comparable to the minimum Dst values of other extreme events from 1859 September and 1921 May. Their minimum Dst estimates are ≈−949 ± 31 nT (Hayakawa et al. 2022) and ≈−907 ± 132 nT (Love et al. 2019b), respectively. There is significant uncertainty in the 1872 min Dst value because it is based on a single station, as is the corresponding value for 1859. Moreover, this single station has a data gap at the time of maximum intensity. The data gap suggests an underestimate of the storm intensity of the Dist $H \leq -834$ nT because it resulted from the placement of a deflector magnet to bring the H-trace back on scale. The Tiflis Dist $H$ (≈−1402 ± 62 nT) indicates the storm was even more intense. In addition, the inferred equatoward boundary of the auroral oval during the storm is 24°.2 ILAT versus 25°.1 ILAT and 27°.1 ILAT for those of 1859 September and 1921 May (Hayakawa et al. 2019b, 2020), respectively, suggesting that the 1872 February storm was even more extreme than the others. The minimum Dst value is estimated to be ≈ −1250 nT for the Chapman–Silverman storm versus ≈ −1200 nT for the Carrington storm in 1859 September (Cliver et al. 2022a), if we apply the empirical correlation between equatoward boundary of the auroral oval and the magnitude of the associated magnetic storm in Figure 3 from Yokoyama et al. (1998). That said, these estimates may be highly uncertain because Figure 3 in Yokoyama et al. (1998) has only three data points for extreme storms < −300 nT. A value of ≈-1250 nT is bracketed by the storm intensity estimates based on the Colaba (Dist $H \leq -834$ nT) and Tiflis (Dist $H \approx -1402 \pm 62$ nT) magnetograms. The fact that there are 13 sites with auroral reports from sites within <|20.0|° MLAT of the magnetic equator for the 1872 February storm versus only one each for the 1859 September and 1921 May events provides additional evidence for the strength of the 1872 storm. At least, this storm was intense enough to cause intense earth currents—likely geomagnetically induced currents—even down to Bombay (10°.0 MLAT) and Khartoum (13°.4 MLAT).

On the basis of the above magnetometer records and auroral observations, it seems clear that the Chapman–Silverman storm ranks among the three largest storms in observational history, comparable in intensity to the Carrington storm of 1859 September and the New York Railway storm of 1921 May. All three of these storms were significantly more intense than the 1989 March storm (minimum Dst=−589 nT; equatoward boundaries of auroral particle precipitation and auroral electric field at 40°.1 and 35° in MLAT, respectively), which was the largest magnetic storm since the International Geophysical Year (Rich & Denig 1992; Yokoyama et al. 1998; WDC Kyoto et al. 2015; Boteler 2019).

## 7. Summary and Discussion

The principal results of this paper are as follows:

1. The extreme storm of 1872 February 4 originated in Secchi's Spot Group #29 that had an area of 461 μsh and a location of S19° E05° in Secchi's drawing (Figure 2) on February 3. From prominence observations by Denza, Secchi, and Tacchini, we infer a solar eruption at ≈9–10 UT on February 3 that we associate with Spot Group #29, yielding a transit time to Earth of ≈ 29 hr based on the storm SC at ≈ 14:27 UT on February 4.

2. The Chapman–Silverman storm was historically large with an inferred minimum hourly Dst* of ≲ −834 nT based on geomagnetic measurements from Colaba. The estimate places the 1872 event among the Carrington event of 1859 September (Hayakawa et al. 2019b, 2020, 2022) and the New York Railway storm of 1921 May (Love et al. 2019b) in the category of extreme magnetic storms (Hayakawa 2020; Cliver et al. 2022a) with minimum Dst* ≲ −800 nT. In terms of auroral visibility, the 1872 storm was pre-eminent, with 13 sites with auroral observations with MLATs <|20°.0|, with the equatormost credible report from Aden at |8°.3| MLAT, versus only one such low-latitude sighting documented for both the 1859 September (|17°.3| MLAT) and 1921 May (|16°.2| MLAT) storms. For a modern comparison, the most intense magnetic storm of the present era, in 1989 March, had a minimum Dst value=−589 nT with overhead aurora at ≈40° ILAT and lowest latitude of auroral observation down to |29°.0| MLAT (Silverman 2006; Boteler 2019).

3. The temporal correspondence of intense auroral activity over a broad range of longitudes (from Barnstaple in England to Shanghai in China) and latitudes (Barnstaple in England to St. Denis in Réunion) was reflected in magnetogram variations from Havana to Greenwich to Bombay.

Cliver et al. (2022b) reported that nearly half (14 of 30) great magnetic storms, defined as those with minimum Dst ⩽ −300 nT, originated in spot groups with area ⩽1000 μsh. The 1872 storm indicates that even extreme (minimum Dst$^*$ ⩽ −900 nT) storms can arise in such medium-sized spot groups. For comparison, the 1921 May storm originated in a spot group of area 1324 μsh on the day of the flare with a corresponding value of 2971 μsh for the 1859 storm (Jones 1955; Hayakawa et al. 2023a).

For good reason, the Carrington event is considered to be an exemplar of an extreme solar-terrestrial event. It ranks high in several categories, including flare size, ICME transit time, and magnetic storm intensity (Cliver & Svalgaard 2004; Cliver & Dietrich 2013). The most likely reason for which the 1872 storm—which originated in a more modest region with an inferred less energetic ICME (based on transit time)—produced an equivalent storm is an unusually strong southward field component in the ICME-driver (e.g., Dungey 1961; Fairfield & Cahill 1966; Tsurutani et al. 1988).

The effects of the 1872 February magnetic storm on technological infrastructure were widely noted and well documented at the time (Preece 1872). For example, Earwaker (1872, p. 323) writes: "The electrical disturbances on the cables in the Mediterranean, and on those between Lisbon and Gibraltar, and Gibraltar and the Guadiana, were also very great. The signals on the land line between London and the Land's End were interrupted for several hours last night by atmospheric currents." Even worse, intense earth currents were reported even down to the low-MLAT regions such as Bombay (10°.0 MLAT) and Khartoum (13°.4 MLAT). In such extreme storms, the rapidly changing magnetic fields of field-aligned and auroral currents can induce currents on the modern power grid at low- and mid-latitudes.

This case study demonstrates the importance of researching past historical records to understand current key space weather issues. Such studies can provide unique scientific details for the extremity of the solar-terrestrial environments, thus underscoring the urgent need for the long-term preservation of these historical scientific data archives (Pevtsov et al. 2019).

## Acknowledgments

We dedicate this paper to the memory of our late friend and colleague Sam Silverman, a distinguished scientist and lawyer, and always good company. We also wish to thank Rachel Rosenblum and Sam Silverman's children for their help in organizing discussions between H.H. and S.M.S. This research was conducted under the financial support of JSPS Grant-in-Aids JP15H05812, JP20K20918, JP20H05643, JP20H01959, JP21K13957, JP20KK0072, JP21H01124, JP21H04492, and JP22K02956. H.H. has been part funded by JSPS Overseas Challenge Program for Young Researchers, the ISEE director's leadership fund for FYs 2021-2023, Young Leader Cultivation (YLC) program of Nagoya University, Tokai Pathways to Global Excellence (Nagoya University) of the Strategic Professional Development Program for Young Researchers (MEXT), and the young researcher units for the advancement of new and undeveloped fields, Institute for Advanced Research, Nagoya University of the Program for Promoting the Enhancement of Research Universities. H.H. acknowledges the International Space Science Institute and the supported International Teams #510 (SEESUP Solar Extreme Events: Setting Up a Paradigm), #475 (Modeling Space Weather And Total Solar Irradiance Over The Past Century), and #417 (Recalibration of the sunspot Number Series). FC and the World Data Center SILSO providing the sunspot number are supported by the Belgian Science Policy Office (BELSPO) via the Solar-Terrestrial Center of Excellence (STCE). D.J.K. was partially supported by NASA Award 80NSSC20K0199, NSF Award Award Number 1933040, and AFOSR Award FA9550-17-1-0258. The work of P.A.B. was partly funded by a grant of the Austrian Science Fund (FWF): P 32958-N. We thank Tadanobu Aoyama, Shota Notsu, and Yuko Ikkatai for their help in accessing Russian auroral reports, Basu (1954), and the Times of India. H.H. thanks Ankush Bhaskar for his advice on interpretations on Moos' data. We thank the Royal Astronomical Society, Osservatorio Astronomico di Roma, and Osservatorio Astrofisico di Catania for allowing us to analyze and reproduce Bernaerts, Secchi, and Tacchini's sunspot drawings. We also thank Biblioteca Pública e Arquivo Regional de Ponta Delgada (Graça Viveiros and Válter Rebelo) and Biblioteca Pública del Estado en Santa Cruz de Tenerife (Javier Machín Godoy) for providing copies of historical newspapers and books. We consulted weather maps in the US Department of Commerce Digital Collection of US Daily Weather Maps. The National Solar Observatory is operated by the Association of Universities for Research in Astronomy, Inc. (AURA), under cooperative agreement with the US National Science Foundation. We thank the Shounji Temple to allow us to analyze and reproduce Figure 10.

## Appendix

Silverman (2008) noted "The Bombay observation [in the 1872 February 6 edition of the Times of India (Figure A1)], like that at Apia [for the 1921 May event; Love et al. 2019b], offers a good description of an aurora." Indeed, the sentence, "After sunset on Sunday, the aurora was slightly visible, and constantly changing color, becoming deep violet when it was most intense—about three o'clock on Monday morning" indicates a closely observed event. Silverman (2008) writes, "We are then left with the question of what to do with the Bombay report, which seems clearly to be credible, and that of Gondokoro. A way out of this puzzle may be a connection with sporadic auroras (see Silverman 2003; for a careful and well documented study of sporadic auroras in China and Japan, see Willis et al. 2007)." As an alternative to this speculative explanation, we offer a less scientific, more prosaic, reason for the descriptive detail in the newspaper account.

There is reason to believe that the article in the Times of India (1872-02-06, p. 2) about the aurora in Bombay is based on a collection of observations that may involve a conflation of the report of the aurora at Jacobabad with telegraph outages at Bombay. For example, the opening question of the article referring to the aurora at Bombay—"Will it surprise our readers to learn that the aurora borealis was plainly visible in Bombay on Sunday night last?"—resembles that from the Times correspondent from Jacobabad (Times of India; 1872-02-15, p. 2)[30] quoted in Chapman (1957b): "Has anyone ever observed the phenomenon of the Aurora Borealis in India?" At Jacobabad, the aurora "displayed light of different hues and colors," and in Bombay it "constantly kept changing color." Both the Jacobabad and Bombay reports refer to the "deep violet" color of the aurora—at Jacobabad during the interval from 0:45 LT (20:11 UT) to 2:15 LT (21:41 UT)—and at Bombay when the aurora "was most intense—about three o'clock on Monday morning." There are other timing discrepancies between the two accounts. At Jacobabad, the aurora was visible from 23:30 LT (18:56 UT) until 2:15 LT (21:41 UT) versus "after sunset (at 17:55 LT; 13:04 UT)" at Bombay when it was "slightly visible" to "until sunrise (at 6:34 LT; 1:43 UT)" on the following morning before which it was "distinctly visible." Given the timing uncertainty in the phrase "after sunset" in the Times article, the first appearance of the aurora in Bombay may be consistent with the 19:18 LT (14:27 UT) timing of the storm sudden commencement at Colaba (sunset at 17:55 LT [13:04 UT] in Bombay on February 4). Neither "after sunset" nor Stacey's "plainly visible" aurora from 20:30 LT (15:39 UT) in Preece (1872) agree with the onset of aurora at Jacobabad, "the sudden change from darkness to light … " at "about half-past 11 p.m. [23:30 LT; 18:56 UT]" that aligns with a sharp positive excursion in the Colaba $H$-trace (Figure 6). The disappearance of the aurora at Jacobabad corresponds to an acceleration of the recovery in the Colaba $H$-trace. The 20:30 LT (15:39 UT) to 4:30 LT (23:39 UT) duration of the aurora at Bombay given by Stacey (Preece 1872) corresponds more closely to the main phase of the storm, whereas the report in the Times of India suggests that the aurora, although no longer observable at Jacobabad, continued past the 6:34 LT (1:43 UT) sunrise in Bombay.[31]

---

[30] While the Jacobabad correspondent's article was not published in the Times until February 15, in it he wrote, "I have subsequently learnt from a telegraph employee that there was a perpetual disturbance of the [magnetic] needle from half-past seven last night till ten this morning...," indicating that the report was dispatched to the Times on February 5.

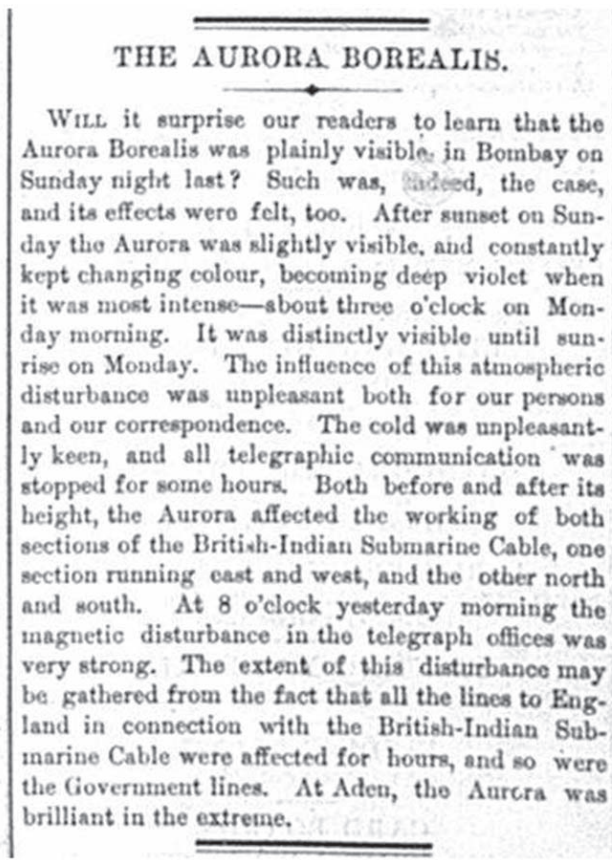

**Figure A1.** Original article Times of India (1872-02-06, p. 2), reproduced with courtesy to © the British Library Board.

---
[31] Secchi (1872b, p. 83); MS OAR 1–4.

**ORCID iDs**

Hisashi Hayakawa https://orcid.org/0000-0001-5370-3365
Edward W. Cliver https://orcid.org/0000-0002-4342-6728
Frédéric Clette https://orcid.org/0000-0002-3343-5153
Yusuke Ebihara https://orcid.org/0000-0002-2293-1557
Shin Toriumi https://orcid.org/0000-0002-1276-2403
Ilaria Ermolli https://orcid.org/0000-0003-2596-9523
Theodosios Chatzistergos https://orcid.org/0000-0002-0335-9831
Kentaro Hattori https://orcid.org/0000-0001-9933-0023
Delores J. Knipp https://orcid.org/0000-0002-2047-5754
Gianna Cauzzi https://orcid.org/0000-0002-6116-7301
Kevin Reardon https://orcid.org/0000-0001-8016-0001
Philippe-A. Bourdin https://orcid.org/0000-0002-6793-601X
Keitaro Matsumoto https://orcid.org/0000-0003-2002-0247
Yoshizumi Miyoshi https://orcid.org/0000-0001-7998-1240
José R. Ribeiro https://orcid.org/0000-0002-4640-1569
Ana P. Correia https://orcid.org/0000-0002-7843-8059
David M. Willis https://orcid.org/0000-0001-9756-8601
Matthew N. Wild https://orcid.org/0000-0002-4028-2300